\documentclass[amsmath,amssymb,
 aps, prd, showkeys,
	nofootinbib,
	floatfix,
        superscriptaddress,]{revtex4-2}

\usepackage{placeins}
\usepackage{graphicx}
\usepackage{dcolumn}
\usepackage{bm}

\usepackage[utf8]{inputenc}
\usepackage{aas_macros}
\pdfoutput=1
\usepackage{hyperref}

\usepackage[caption=false]{subfig}
\hypersetup{ 
			colorlinks=true,
			pdfauthor={Kyle Boone},
			pdftitle={WST TBD},
			citecolor=blue
			}

\usepackage{xcolor}
\usepackage{multirow}
\usepackage{array} 
\usepackage{ctable}
\usepackage[normalem]{ulem}

\newcommand{\lp}{\left(}
\newcommand{\rp}{\right)}

\begin{document}

\title{Constraining Power of Wavelet vs. Power Spectrum Statistics for CMB Lensing and Weak Lensing with Learned Binning}

\author{Kyle Boone}
\email{kboone@g.harvard.edu}
\affiliation{Department of Physics, Harvard University, Cambridge, MA 02138, USA}

\author{Georgios Valogiannis}
\email{gvalogiannis@uchicago.edu}
\affiliation{Department of Astronomy and Astrophysics, University of
Chicago, Chicago, IL 60637, USA}
\affiliation{Kavli Institute for Cosmological Physics, Chicago, IL, 60637, USA}

\author{Marco Gatti}
\email{mgatti@uchicago.edu}
\affiliation{Kavli Institute for Cosmological Physics, Chicago, IL, 60637, USA}

\author{Cora Dvorkin}
\email{cdvorkin@g.harvard.edu}
\affiliation{Department of Physics, Harvard University, Cambridge, MA 02138, USA}

\date{\today}
             
\begin{abstract}
We present forecasts for constraints on the matter density ($\Omega_m$) and the amplitude of matter density fluctuations at 8h$^{-1}$Mpc ($\sigma_8$) from CMB lensing convergence ($\kappa_\text{CMB}$) maps and galaxy weak lensing convergence ($\kappa_\text{WL}$) maps.
For $\kappa_\text{CMB}$ auto statistics, we compare the angular power spectra ($C_\ell$'s) to the wavelet scattering transform (WST) coefficients.
For $\kappa_\text{CMB}\times\kappa_\text{WL}$ statistics, we compare the cross angular power spectra to wavelet phase harmonics (WPH). This work also serves as the first application of WST and WPH to these probes. 
For $\kappa_\text{CMB}$, we find that WST and $C_\ell$'s yield similar constraints in forecasts for all surveys considered in this work. When $\kappa_\text{CMB}$ is crossed with $\kappa_\text{WL}$ projected from \textit{Euclid} Data Release 2 (DR2), we find that WPH outperforms cross-$C_\ell$'s by factors between $2.2$ and $3.4$ for individual parameter constraints.
To compare these different summary statistics, we develop a novel learned binning approach. This method compresses summary statistics while maintaining interpretability. We find this leads to improved constraints compared to more naive binning schemes for our wavelet-based statistics, but not for $C_\ell$'s. By learning the binning and measuring constraints on distinct data sets, our method is robust to overfitting by construction.
\end{abstract}

\maketitle
\twocolumngrid

\section{Introduction}
\label{sec:introduction}

Gravitational lensing of photons occurs when light is deflected by a mass distribution between the source and the observer \citep{Einstein_Lensing}.
By observing distortions in the shapes of numerous galaxies from weak lensing (hereafter WL), statistical methods can be used to obtain integrated line-of-sight mass maps \citep[e.g.][]{WL_Example}.
Current and upcoming surveys, such as the \textit{Euclid} mission \citep{Euclid}, the Dark Energy Survey \citep[DES,][]{DES}, the Kilo-Degree Survey \citep[KIDS,][]{Kids}, the Hyper Suprime-Cam \citep[HSC,][]{HSC}, and the {\it Vera C. Rubin} Observatory's Legacy Survey of Space and Time \citep[LSST,][]{LSST}, will measure galaxy shapes, enabling measurements of these integrated mass maps.
Additionally, by observing the lensing of the cosmic microwave background (hereafter CMBL), one can obtain mass maps integrated all the way back to the surface of last scattering \citep[e.g.][]{CMBL_Ex}.
Examples of surveys that perform this measurement include \textit{Planck} \citep{Planck}, the Atacama Cosmology Telescope (ACT) \citep{ACT}, the South Pole Telescope (SPT) \citep{SPT}, and the \textit{Simons} Observatory (SO) \citep{SO}.
Lensing-convergence maps from these surveys are used to constrain cosmology by comparing measured summary statistics with predictions from analytic theories or forward-modeled simulations, within either a likelihood-based or simulation-based inference framework.

If the convergence maps were Gaussian random fields, all of their information would be captured by two-point statistics. However, due to non-linear gravitational evolution, the large-scale structure of the universe has non-Gaussianities that become more pronounced at low redshift. 
Since two-point statistics fail to capture non-Gaussian information, other summary statistics are desirable.
Numerous summary statistics have been used to try to extract non-Gaussian information, including the wavelet scattering transform \citep[e.g.][]{Cosmo_WST, PhysRevD.106.103509,PhysRevD.105.103534,PhysRevD.109.103503,DES_WST, HSC_WST} and wavelet phase harmonics \citep[e.g.][]{DES_WST}. Other approaches include higher-order N-point statistics \citep[e.g.][]{10.1093/mnras/stv961, Philcox:2021hbm, Ivanov:2021kcd, Chudaykin:2025aux,sunao2025}, density-split clustering \citep{Paillas:2023cpk}, topological descriptors such as Minkowski functionals \citep{2025JCAP...05..064L,prat2025, Armijo_2025}, and machine-learning-based summaries like convolutional neural networks \citep{Fluri_2019, jeffrey2025}.

In this work, we compare the information content of selected summary statistics applied to the lensing of the cosmic microwave background and weak lensing fields using Fisher forecasts.
Each summary statistic compresses the information in a field and typically depends on a handful of hyperparameters (such as the range of scales in a power spectrum or the number of different opening angles in three-point statistics).
Hyperparameter choices change the effective scales and the dimensionality of the summary statistics; this can impact the stability of the covariance inversion in Gaussian-likelihood pipelines and, for simulation-based inference pipelines, inflate model complexity and training variance. Constraints become then highly dependent on hyperparameters. We address this by proposing and developing a novel learned binning approach, which produces compact and informative summaries.

In our approach, each summary statistic has access to all scales present in our maps.
The statistics are then compressed via binning to the same size while maximizing the information content (the determinant of the Fisher matrix).
By learning our binning and providing parameter constraints on distinct data sets, our method is robust to overfitting and yields a compressed statistic without sacrificing interpretability.

The rest of the paper is organized as follows: in Section \ref{sec:data} we introduce the simulations used in our analysis and explain how to use them to construct convergence maps.
In Section \ref{sec:sum}, we introduce the summary statistics used.
In Section \ref{sec:analysis}, we present an overview of our learned binning approach, which is then applied to simulated data in Section \ref{sec:results}. 
Finally, we conclude in Section \ref{sec:conclusion}. 
All parameter constraints are presented in Appendix \ref{app:all_res}, and their convergence is tested in Appendix \ref{app:conv_move}.
We test the convergence of our derivatives in Appendix \ref{app:conv_ders}.
Additional details on the learned binning approach are given in Appendix \ref{app:bin}.
We verify the Gaussianity of our summary statistics required for a reliable Fisher forecast in Appendix \ref{app:gauss}.

\section{Simulated Observables}
\label{sec:data}

We use weak gravitational lensing convergence as our simulated observable.
Weak gravitational lensing distorts distant images and can be summarized by the lensing potential $\psi$, which is an integral of the gravitational potential $\Phi$ along the line of sight:

\begin{equation}
    \psi(\vec{\theta},D_S) = \frac{2}{c^2D_S}\int_0^{D_S}dD_L \frac{D_{LS}}{D_L}\Phi(D_L\vec{\theta},D_L),
\end{equation}
where $D_L$, $D_S$, and $D_{LS}$ are the angular distances from the observer to the lens, the observer to the source, and the lens to the source, respectively, $c$ is the speed of light, and $\vec{\theta}$ is the angular position on the sky.
From the lensing potential, we can then construct the convergence map $\kappa$ as:

\begin{equation}
    \kappa(\vec{\theta},D_S) \equiv \frac{1}{2}\nabla^2_\theta \psi(\vec{\theta},D_S). 
\end{equation}

In this work we exclusively use convergence maps integrated out to the cosmic microwave background (CMB) ($\kappa_\text{CMB}(\vec{\theta})$) and convergence maps integrated out to redshift ranges of galaxy surveys ($\kappa_\text{WL}(\vec{\theta})$).
These are defined in the following way:

\begin{gather}
    \kappa_\text{CMB}(\vec{\theta})\equiv\kappa(\vec{\theta}, D_{S,\text{CMB}}), \\
    \kappa_\text{WL}(\vec{\theta})\equiv\kappa(\vec{\theta}, D_{S,\text{WL}}),
\end{gather}
where $D_{S,\text{CMB}}$ and $D_{S,\text{WL}}$ are the angular distances to the CMB and redshift bin of galaxies, respectively.
We now turn to the simulations used to generate convergence maps.

\subsection{Simulations: \texttt{ULAGAM} Suite}

The analysis in this work uses the \texttt{ULAGAM} suite of N-body simulations \citep{Ulagam}. The suite consists of 2000 simulations at a fiducial cosmology, plus sets of 100 simulations where one parameter is varied at a time in a stepwise fashion for Fisher forecast purposes.

The simulations were run using \texttt{PKDGRAV3} \citep{Potter2016}, and are performed in periodic cubic boxes of size $L = 1\,h^{-1}\mathrm{Gpc} \approx 1.5\,\mathrm{Gpc}$, initialized at redshift $z = 127$ and evolved with $512^3$ dark matter particles. The initial conditions for all simulations were sourced from the Quijote suite \citep{Villaescusa-Navarro2020}. The fiducial cosmological model is $\Lambda$ Cold Dark Matter ($\Lambda$CDM) and assumes the following parameters: matter density $\Omega_m = 0.317$, baryon density $\Omega_b = 0.049$, reduced Hubble constant $h = 0.671$, spectral tilt $n_s = 0.962$, and amplitude of matter density fluctuations at 8h$^{-1}$ Mpc $\sigma_8 = 0.834$. 

Our analysis focuses on forecasting constraints on $\Omega_m$ and $\sigma_8$, the $\Lambda$CDM parameters determined by weak lensing. 
To evaluate the derivatives necessary for a Fisher forecast, we use simulations where $\Omega_m$ is shifted by $\pm0.01$ and where $\sigma_8$ is shifted by $\pm0.015$.
We verify in Appendix \ref{app:conv_ders} that these parameter shifts are small enough for derivative estimates to converge.
We use 1000 realizations for the fiducial cosmology, and 100 realizations each for variations in $\sigma_8$ and $\Omega_m$. \texttt{PKDGRAV3} automatically generates particle lightcones in the form of \texttt{HEALPix}\footnote{\url{http://healpix.sourceforge.net}} pixel maps \citep{HEALPix} at varying redshifts. These maps are provided at $\texttt{NSIDE}=1024$, corresponding to a pixel size of roughly 3.44 arcminutes. The simulation box is tiled and repeated as needed to construct sufficiently large volumes for building full-sky lightcones up to a given redshift. We convert the particle maps into overdensity maps and use them to generate weak-lensing convergence maps, which we discuss in the next subsection. 

\subsection{CMB Lensing Convergence}

\begin{table}
    \centering
    \begin{tabular}{|l|c|c|}
        \hline
 \multicolumn{3}{|c|}{Survey Footprint Areas (deg$^2$)} \\
 \hline \hline\hline\textbf{Survey} & Survey Area & Overlap with \textit{Euclid} DR2\\ 
        \hline \hline
        
        ACT & 15472 & 2903\\ \hline 

        \textit{Planck} & 24915 & 7290 \\ \hline

        SPT & 4001 & 2415 \\ \hline

        SO & 27229 & 3075 \\ \hline
    \end{tabular}
    \caption{Coverage area for all CMB surveys considered in this work and their overlap area with \textit{Euclid} DR2.}
    \label{tab:survey_areas}
\end{table}

\begin{figure}
  \includegraphics[width=0.45\textwidth]{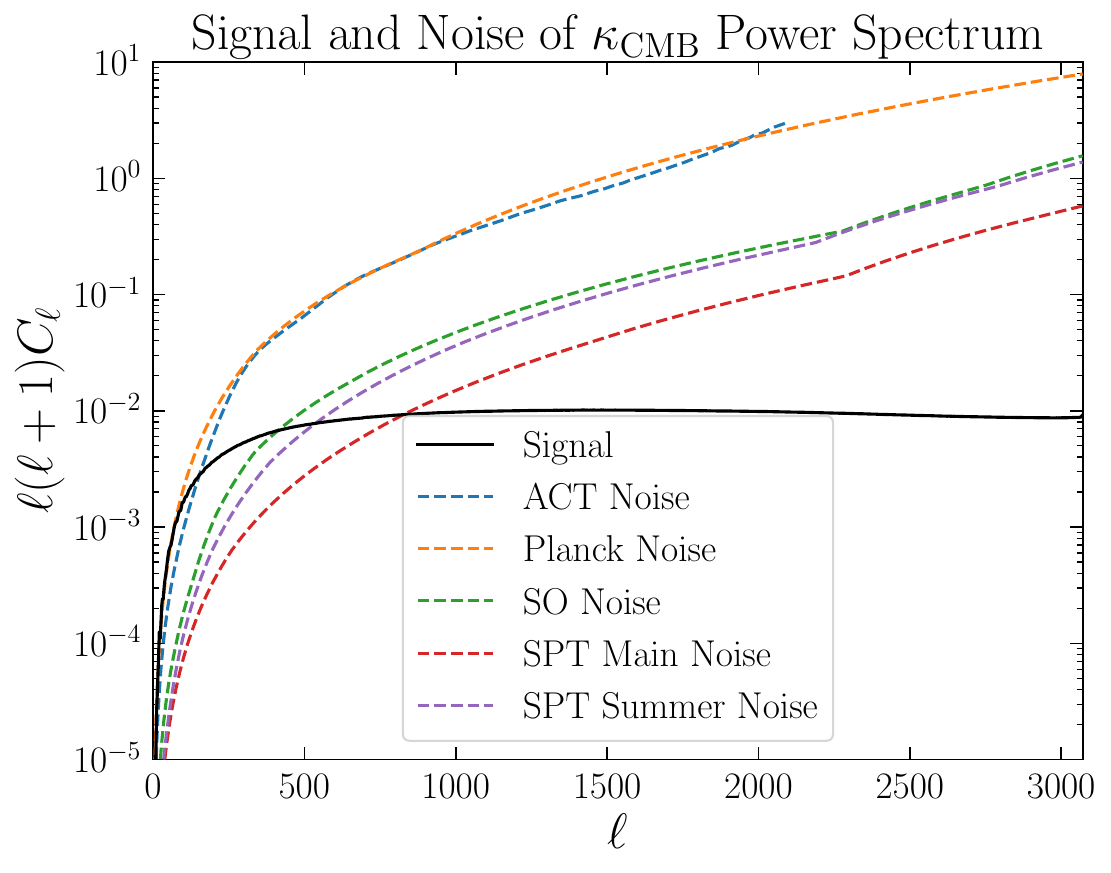}
  \caption{We show the average noiseless auto power spectrum of $\kappa_\text{CMB}$ (black) in comparison to noise power spectra from all CMB surveys used (dashed lines). For ACT, we only plot until $\ell=2090$ as this is the limit of our noise power spectrum data. For SPT, we plot the noise power spectra from the main and summer fields separately. 
  The cusps (for instance, at $\ell\approx2200$ for SPT) occur because our noise power spectra are linear interpolations between specified points.}
  \label{fig:signal_noise}
\end{figure}

\begin{figure}
  \includegraphics[width=0.45\textwidth]{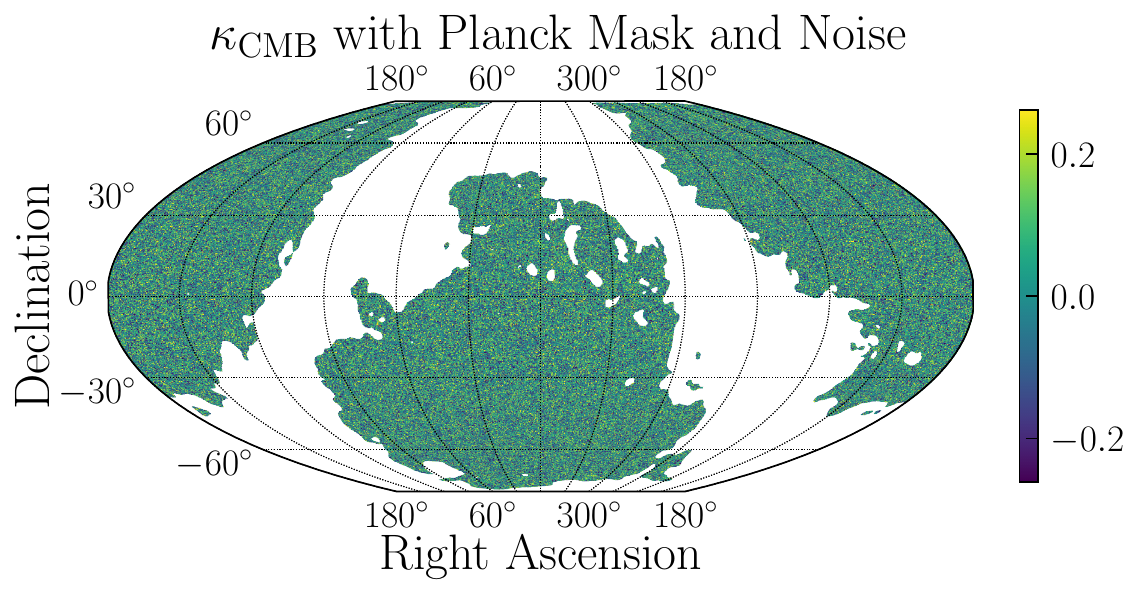}
  \caption{We show $\kappa_\text{CMB}$ with \textit{Planck} levels of noise on \textit{Planck}'s footprint.}
  \label{fig:ex_map}
\end{figure}

We follow the procedure described in Section $3.1$ of \citet{Convergence} to create noiseless, full-sky $\kappa_\text{CMB}$ maps for each of our simulations. In particular, we use the Born approximation to integrate the overdensity maps up to redshift $z = 3.5$. The CMB lensing power spectrum is then computed for the range $3.5 < z < 1089$, and a Gaussian realization based on this spectrum is generated and added to obtain the full $\kappa_\text{CMB}$ map.

We generate mock survey $\kappa_\text{CMB}$ maps by adding Gaussian noise and footprint masks to the full-sky, noiseless convergence maps, simulating various experiments. Specifically, we produce maps for \textit{Planck}, ACT Data Release 6 (DR6), SPT main and summer fields, and SO. The noise is modeled as purely Gaussian and generated from the corresponding noise power spectra, taken from \citet{SO} and \citet{SPT_noise}.
For ACT, our noise power spectra only go to $\ell_\text{max}=2090$, so we set all signal beyond this to zero.
Signal and noise power spectra for $\kappa_\text{CMB}$ are shown for all surveys used in Figure \ref{fig:signal_noise}.

For \textit{Planck}, we further apply a Gaussian smoothing beam with a full width half maximum of 10 arcminutes after adding noise. This mimics the instrument's beam. We ignore the beams of the other experiments because they are smaller than the pixel size of our maps. Survey masks are applied to each simulated map to match the respective sky coverage. For SPT, we consider the union of the main and summer fields, each region with its corresponding noise power spectrum applied.
The coverage area of all surveys can be found in Table \ref{tab:survey_areas}.
An example $\kappa_\text{CMB}$ for \textit{Planck} levels of noise is shown in Figure \ref{fig:ex_map}.

We note that this procedure is idealized and produces simplified CMB lensing maps with Gaussian noise and no additional contaminants. These maps are sufficient for the purpose of this work, namely, a forecast-level assessment of the intrinsic non-Gaussian information in the lensing field, but they are not suitable for direct comparison with real data. In practice, CMB lensing is not measured directly but reconstructed from multi-frequency temperature and polarization maps using, for example, quadratic estimators \citep{Hu_2002}, after applying standard data-cleaning steps such as frequency filtering and survey-mask corrections. Since lensing is a small signal superimposed on the primary CMB anisotropies, the reconstructed maps are contaminated by the CMB itself and by reconstruction effects that must be calibrated using simulations. Realistic CMB lensing mocks, therefore, require simulating unlensed CMB maps, lensing them with a full-sky convergence field, adding realistic anisotropic and non-Gaussian noise, and reconstructing the lensing field using the same estimators and corrections applied to the data \citep{Liu_2016}. The resulting maps naturally include non-Gaussian reconstruction noise and higher-order bias terms. Our aim here is more limited: to quantify, at the forecast level, the amount of intrinsic non-Gaussian information in the lensing field itself. For this purpose, modelling the noise as Gaussian is sufficient to capture the correct order-of-magnitude behavior, as we do not attempt to reproduce the detailed statistics of reconstructed CMB lensing maps.

\subsection{Weak Galaxy Lensing Convergence}

To create mock $\kappa_\text{WL}$ maps, we use a simplified version of the pipeline described in \citet{DES_WST}, tailored to resemble \textit{Euclid} DR2 data. First, we generate noiseless convergence maps as a function of redshift from the density field using the Born approximation. 

We assume that the \textit{Euclid} DR2 weak-lensing source sample is divided into six tomographic bins of roughly equal number density. 
For simplicity, we use only two tomographic bins: the lowest-redshift bin and the highest-redshift bin.
The first (last) tomographic bin has a mean redshift of 0.33 (1.37) and a standard deviation of 0.09 (0.16). 
For each bin, we produce a convergence map by summing the redshift shells weighted by the corresponding redshift distribution. 
We add Gaussian shape noise, assuming an ellipticity dispersion of $\sigma_e = 0.26$ and a galaxy number density of 30 galaxies per arcmin$^2$, uniformly distributed across the tomographic bins \citep{Euclid}. 
Finally, we mask the mock $\kappa_\text{WL}$ maps by applying the \textit{Euclid} DR2 footprint mask. 
For all $\kappa_\text{CMB}\times\kappa_\text{WL}$ statistics, we apply both the \textit{Euclid} and CMB-survey masks. The overlap coverage area for all surveys can be found in Table \ref{tab:survey_areas}.

\section{Summary Statistics}
\label{sec:sum}

For $\kappa_\text{CMB}$ maps, we quantify information using the angular power spectrum ($C_\ell$) and the wavelet scattering transform (WST).
For $\kappa_\text{CMB}\times\kappa_\text{WL}$, we generalize these statistics to the cross angular power spectrum and wavelet phase harmonics (WPH).

\subsection{Power Spectrum}

For a function $A(\Omega)$ on the sphere, $a^A_{\ell m}$ values are defined as:

\begin{equation}
    a^A_{\ell m} = \int A(\Omega)Y_{\ell,m}^*(\Omega)\,\,d\Omega ,
\end{equation}
with $Y_{\ell,m}$ denoting the spherical harmonics and $\Omega$ the solid angle.
From the $a_{\ell m}$'s, we define the cross angular power spectra as:

\begin{equation}
    C^{AB}_\ell = \frac{1}{2\ell+1}\sum_{m=-\ell}^\ell a^A_{\ell,m} a^{B*}_{\ell,m}.
\end{equation}

The auto angular power spectra are defined with $A=B$.
From here on, we will omit superscripts as the context will make it clear which power spectrum we are referring to.
The $C_\ell$'s are all computed on the masked fields described in Section \ref{sec:data}.

The $C_\ell$'s are able to capture all Gaussian information about a field, but miss any information contained in non-Gaussianities.
An example of a summary statistic for a single map that can capture non-Gaussian information is the WST, which we turn to next.

\subsection{Wavelet Scattering Transform}

The WST was initially proposed in \citet{Mallat} and then used in cosmology to reduce noise in simulations \citep{Romeo_2003} and extract information from weak lensing maps \citep{Cosmo_WST} and galaxy clustering \citep{PhysRevD.105.103534,PhysRevD.106.103509,PhysRevD.109.103503}.
WST consists of convolving the initial map with a directional, multi-scaled wavelet and taking the modulus of the result.
Multiple successive convolutions can be performed to probe non-Gaussianities.
An average over all pixels is used to convert any map into a scattering coefficient, the building block of this summary statistic.

More formally, starting with an input field, $A_0$, one can construct:

\begin{equation}
    A_{n+1} = |A_n \star \psi|.
\end{equation}

In the above, $\psi$ is the localized wavelet.
In general, $\psi$ will have a scale and a directional dependence, represented here by the integer-valued parameters $j$ and $l$, respectively.
In this case, one can obtain the first three orders of scattering coefficients as:

\begin{gather}
    S_0=\langle A_0\rangle,\\
    S_1(j_1, l_1) = \langle |A_0 \star \psi_{j_1, l_1}|\rangle, \\
    S_2(j_1, l_1, j_2, l_2) = \langle ||A_0 \star \psi_{j_1, l_1}|\star\psi_{j_2, l_2}|\rangle,
\end{gather}
where $\langle . . .\rangle$ is a spatial average.
As in other works \citep[e.g.][]{DES_WST, HSC_WST}, we only use $S_1$ and $S_2$ coefficients.
If the overlap of wavelets $\psi_{j,l}$ and $\psi_{j,l +1}$ is small and we have statistical isotropy, we can average over orientations without losing information: $S_1(j_1, l_1)\rightarrow S_1(j_1)$ and $S_2(j_1, l_1, j_2, l_2) \rightarrow S_2(j_1, j_2, |l_1 - l_2|)$.
In this work, we assume isotropy for all relevant summary statistics and average over opening angles. 
Additionally, not all scale combinations contribute to $S_2$.
As shown in \citet{WST_guide}, only $S_2$ coefficients with $j_2>j_1$ are significant.

In this work, we use the Morlet wavelets, given by:

\begin{gather}
    \psi_{j,l}(\vec{x}) = \frac{1}{\sigma}\exp\lp-\frac{x^2}{2\sigma^2}\rp [\exp(i\vec{k}_0\cdot\vec{x})-\beta],\label{eq:Morlet}\\
    \tilde{\psi}_{j,l}(\vec{k}) = \exp\lp -\frac{(\vec{k}-\vec{k}_0)^2\sigma^2}{2}\rp - \beta\exp\lp -\frac{k^2\sigma^2}{2}\rp,
\end{gather}
where $\sigma=0.8\times 2^j$, $k_0 = 0.75\pi\times 2^{-j}$, $\text{arg}(\vec{k}_0) = \theta$, with $\theta = (\frac{L}{2}-1-l)\pi/L$. 
The counterterm $\beta=\exp\lp -\frac{k_0^2\sigma^2}{2}\rp$ is introduced so that $\tilde\psi(0)=0$.
The $j$ index will take integer values between $0$ and $J-1$, where $J$ controls the number of scales and is at most $\log_2(\text{Image Size})-1$.
This keeps the wavelet peaks in Fourier space roughly between the fundamental frequency and the Nyquist frequency of the image.
The $l$ values range between $0$ and $L-1$, where $L$ specifies the number of different angles to probe.
While Morlet wavelets traditionally have the above dyadic spacing, we also experiment with finer spacing.
Details on this will be given explicitly before any results are shown.

To apply the wavelet scattering transform, the input image must be two-dimensional and flat.
To achieve this with spherical data, we split the data into 768 disjoint patches, which are the \texttt{HEALPix} pixels at $\texttt{NSIDE} = 8$.
The center of each of these patches is rotated to the north pole of the sphere.
From here, gnomonic projection is used to project the high-resolution patch onto a $256\times 256$ grid \citep[as in][]{DES_WST}, which sets our maximum $J$ value at $J=7$.
Due to spherical distortions, not all \texttt{HEALPix} pixels are mapped one-to-one onto a pixel in the final grid.
We do not attempt to correct for this effect in this work.
After obtaining scattering coefficients for each patch, we average them to obtain scattering coefficients for the entire map.

Since our data is masked, we use the convention to only use a patch if at least 20\% of its \texttt{HEALPix} pixels are unmasked.
The unmasked fraction for a patch is used to weight all of its scattering coefficients prior to averaging across patches. 
We use the wavelet scattering repository\footnote{\url{https://github.com/SihaoCheng/scattering_transform}} for calculating scattering coefficients after modifying it to allow for non-dyadic spacing.

\subsection{Wavelet Phase Harmonics}

Wavelet phase harmonics (WPH) is a generalization of some of the concepts of WST to two maps, which we will denote as $A$ and $B$.
Since WPH uses two maps instead of the one used for WST, it allows for the study of cross-correlations. 
Unlike WST, each map is only ever convolved with one wavelet, which we denote as:

\begin{equation}
    A_{j,l} = A\star \psi_{j,l}.
\end{equation}
As in WST, $\psi_{j,l}$ is a localized wavelet characterized by a scale parameter $j$ and a directional parameter $l$.
In general, $A_{j,l}$ will be complex-valued.
We apply a phase harmonic to this, defined as:

\begin{equation}
    \text{PH}(r\exp(i\theta), q) = r\exp(iq\theta),
\end{equation}
with $r$, $\theta$, and $q$ real-valued numbers. 

From the equation above, we can define WPH coefficients in general. 
As in WST, because we are assuming isotropy, the only angular dependence will be in $\Delta l\equiv |l_2-l_1|$, and we can write our WPH coefficients as:

\begin{align}
    \text{WPH}_{q_1q_2}(A,B,j_1,j_2,\Delta l) &= \nonumber\\ \langle \text{PH}(A_{j_1,l},q_1) &\text{PH}(B_{j_2,l+\Delta l},q_2)\rangle.
\end{align}

The WPH coefficients that we use are:

\begin{gather}
    S_{00}(A,B,j) = \langle |A_{j,l}||B_{j,l}|\rangle,\\
    S_{01}(A,B,j) = \langle |A_{j,l}|B_{j,l}\rangle,\\
    S_{10}(A,B,j) = \langle A_{j,l}|B_{j,l}|\rangle,\\
    S_{11}(A,B,j) = \langle A_{j,l} B_{j,l}\rangle,\\
    C_{01}(A,B,j_1,j_2,\Delta l) = \langle |A_{j_1,l}|B_{j_2,l+\Delta l}\rangle,\\
    C_{10}(A,B,j_1,j_2,\Delta l) = \langle A_{j_1,l}|B_{j_2,l+\Delta l}|\rangle,
\end{gather}
with $j_1<j_2$ for $C_{01}$ and $j_2<j_1$ for $C_{1,0}$.

These statistics are chosen following \citet{DES_WST}, with the only change being that we allow for more opening angles in $C_{01}$ and $C_{10}$ statistics.
For simplicity, we only consider the real component of the final statistics.

In this work, we use the package PyWPH \citep{PyWPH} to apply WPH and use the bump-steerable wavelets, which are given by the following mother wavelet:

\begin{equation}
    \psi(\vec{k}) = 0.693\exp\lp\frac{-(k-\xi_0)^2}{\xi_0^2-(k-\xi_0)^2}\rp \cos^3(\text{arg}(\vec{k})).
\end{equation}
The equation above is for Fourier wavenumbers in the range $0<k<2\xi_0$ and $|\text{arg}(k)|<\pi/2$; for all other values $\psi =0$.
The normalization and the central frequency $\xi_0=0.85\pi$ are set as in \citet{PyWPH}.
All other wavelets are generated by dilating or rotating the mother wavelet.
In real space:

\begin{equation}
    \psi_{j,l}(\vec{x}) = 2^{-j} \psi\lp 2^{-j}\text{Rot}_{-\pi l/L}(\vec{x})\rp.
\end{equation}
In the above, ``Rot" means to rotate the vector by the subscripted angle. We probe $L=4$ different angles.

To apply WPH, the input images must be two-dimensional and flat, as for WST.
We follow the same procedure as we used for WST to project patches of the sphere onto $256\times 256$ grids.
As in the case of WST, we only use a patch if at least 20\% of its \texttt{HEALPix} pixels are unmasked.
The unmasked fraction for a patch is used to weight the WPH coefficients from each patch before averaging across patches to obtain the WPH coefficients for the entire map.

\section{Analysis}
\label{sec:analysis}

We use the Fisher information approach to compare the constraining power of our different summary statistics.
In doing so, we assume that our summary statistics are Gaussian-distributed, which we test and confirm in Appendix \ref{app:gauss}.
To fairly compare different summary statistics, we compress them with a novel learned binning approach.
This approach compresses the statistics to a set number of bins while maximizing the determinant of the Fisher matrix.

\subsection{Fisher Forecast}

To generate our covariance matrix $C$, we use 1000 fiducial simulations.
Derivatives for the Fisher forecast are estimated numerically using $100$ realizations per parameter shift.
We concatenate our derivative vectors into a matrix $D$, so we can write: 

\begin{equation}
\label{eq:Fish}
    F = D^TC^{-1}D.
\end{equation}

To unbias our inverse covariance, we apply the Hartlap correction \citep{Hartlap}; however, this correction alone does not yield an unbiased Fisher matrix.
Noise in derivative estimates causes the Fisher matrix to systematically overestimate the constraining power of a summary statistic \citep{Combined_Fisher}.

To address this, \citet{Combined_Fisher} introduce a compressed Fisher matrix, $F^\text{comp}$. 
This Fisher matrix is constructed as in Eq.\ \ref{eq:Fish}, but using summary statistics compressed with the MOPED \citep{Heavens2000} algorithm.
Perfect compression of the summary statistics would lead to no information loss, but the compression relies on the matrices $D$ and $C$.
\citet{Combined_Fisher} show that $F^\text{comp}$ will systematically underestimate the constraining power if $D$ and $C$ are estimated from simulations via Monte Carlo methods before compressing the statistics.
Taking the geometric mean of $F$ and $F^\text{comp}$ gives the combined Fisher matrix, $F^\text{Combined}$, in which the opposing biases of the two inputs cancel, yielding unbiased constraints.
It is this combined Fisher matrix and the resulting parameter covariance matrix, $(F^\text{Combined})^{-1}$, which we use for all shown constraints.

The construction of $F^\text{comp}$ requires a random splitting of the derivatives, leading to stochasticity in $F^\text{Combined}$.
To reduce shot noise, we generate $100,000$ realizations of the parameter covariance matrix with different random splits of the derivatives.
Our final parameter covariance matrix is the average of the middle $90,000$ realizations ordered by determinant.
We cut off the most extreme $5,000$ realizations on either end to reduce the impact of extreme outliers, which can occur in this method.
This averaging reduces shot noise while still leaving us sensitive to systematic convergence issues, which we test for.

Before constructing the combined Fisher matrix, we need to compress our summary statistics to obtain an invertible data covariance matrix.
Compression is necessary since the finite number of realizations used to evaluate the covariance matrix places an upper bound on its rank.
Since all of our summary statistics have scale dependence, in this work we implement a novel learned binning (over scale) approach to compress our statistics, while retaining interpretability. We explain our method below.

\subsection{Learned Binning Compression}

\begin{figure}
  \includegraphics[width=0.45\textwidth]{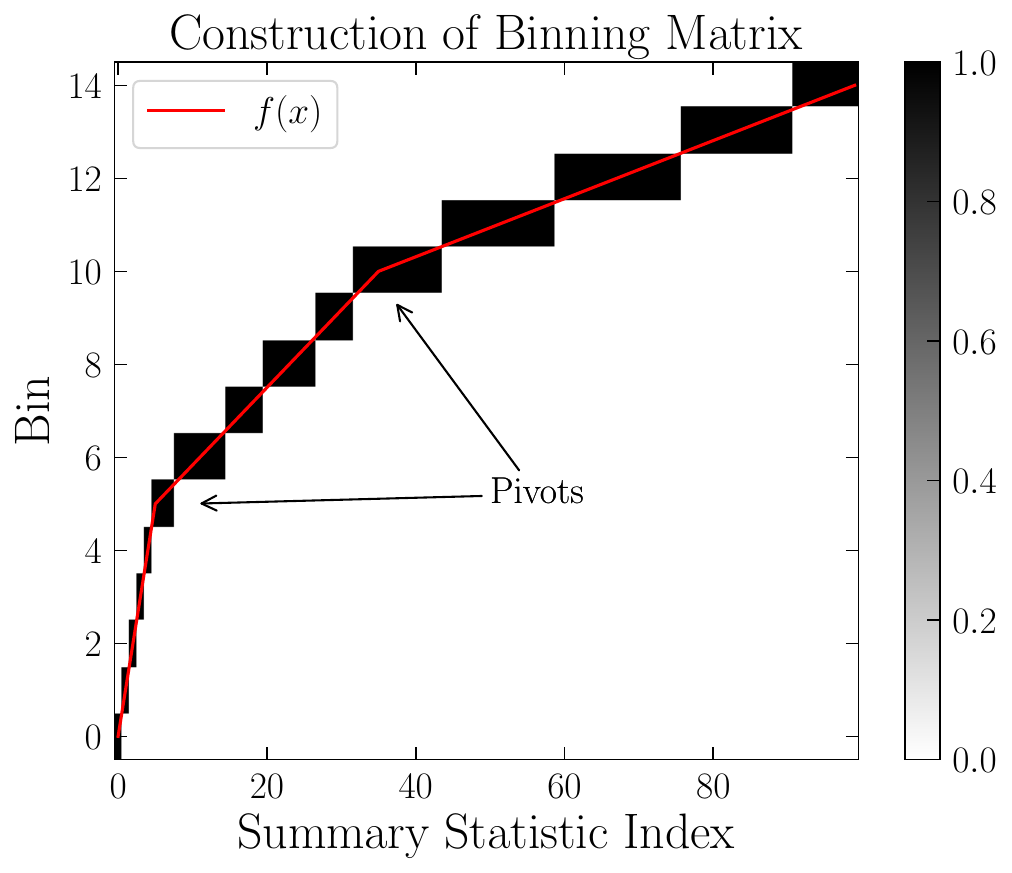}
  \caption{The construction of a binning matrix from a function $f(x)$. Our function is a linear interpolation between labeled pivots. This binning matrix compresses a summary statistic of length $100$ to $15$ bins.}
  \label{fig:bin_construction}
\end{figure}

\begin{figure}
  \includegraphics[width=0.45\textwidth]{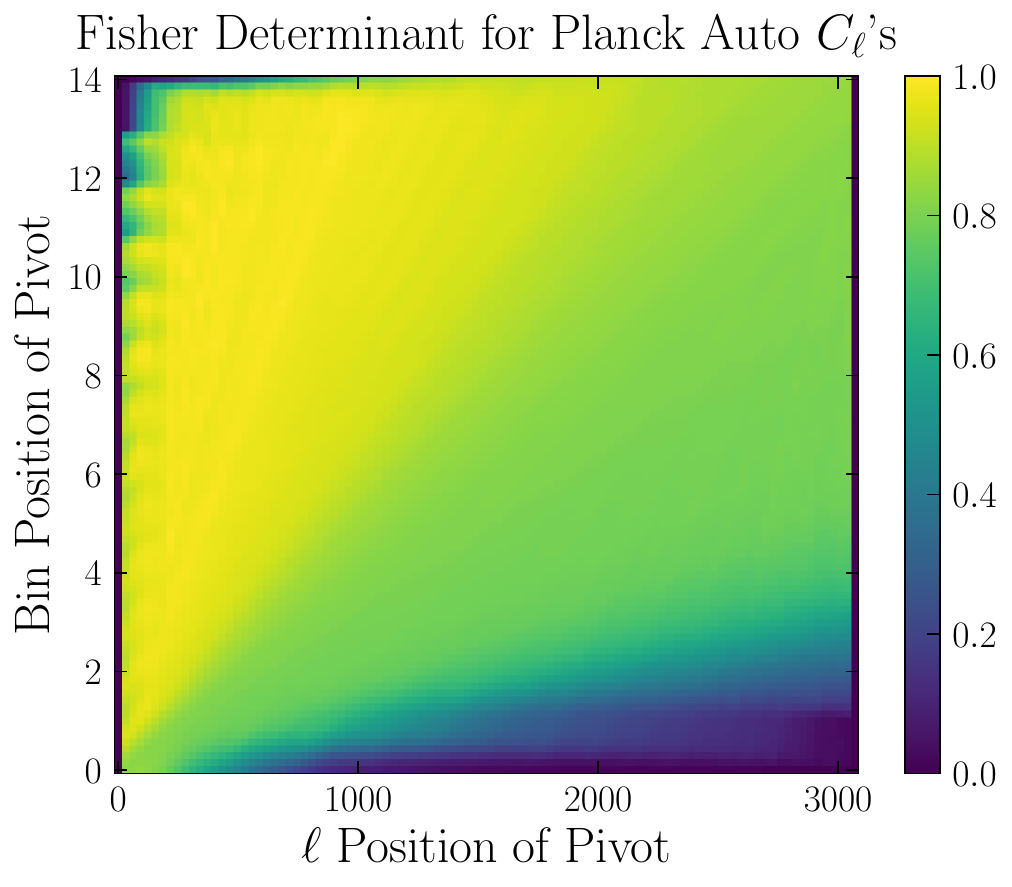}
  \caption{The determinant (normalized to have a maximum of $1$) of the Fisher matrix from Eq.\ \ref{eq:Fish} over pivot space for \textit{Planck} $\kappa_\text{CMB}$ $C_\ell$'s. Only one pivot is used to enable visualization via a heatmap, so the axes denote the $\ell$ and bin positions of the pivot.}
  \label{fig:heatmap}
\end{figure}

In this section, the initial summary statistic will be $M$-dimensional, and it will be compressed to $N$ dimensions, with $N<M$.
Any binning scheme can be viewed as a particular linear transformation within $\mathbb{R}^{N\times M}$.
We will refer to the general binning matrix as $B$. To compress a data vector $\vec{d}$, we perform $B\vec{d}$.

Initially, with $D\in \mathbb{R}^{M\times 2}$ (we have $2$ cosmological parameters) and $C\in \mathbb{R}^{M\times M}$, one would be unable to calculate the Fisher matrix as in Eq.\ \ref{eq:Fish} if $M$ exceeds the number of simulations.
With a binning matrix, one would instead perform:

\begin{equation}
    F = (BD)^T(BCB^T)^{-1}BD.
\end{equation}
In this way, we are now inverting an $N\times N$ matrix instead of an $M\times M$ matrix.
Therefore, we want the value of $N$ to be much smaller than the number of fiducial simulations.
Everything so far simply applies to general $N\times M$ linear transformations beyond just binning matrices.
Compared with general linear transformations, binning is more intuitive and interpretable while having fewer free parameters, thereby reducing the risk of overfitting.

To enable low-dimensional parameterizations of binning matrices,  we use a function $f(x):[1,M]\rightarrow [1,N]$ to specify the positions of nonzero matrix entries.
Specifically, the $m^\text{th}$ column of the binning matrix has a single nonzero entry equal to $1$ at the integer index closest to $f(m)$.
In this work, $f(x)$ is a linear interpolation between $(1,1)$, learned pivots, and $(M,N)$.
Further details on the conditions that $f(x)$ must satisfy can be found in Appendix \ref{app:bin}.
An example of such a construction is shown in Figure \ref{fig:bin_construction}.
We will often refer to the summary statistic index and bin number of the pivot as the ``$x$" or ``$y$" components of the pivot.

To compress summary statistics, we first pick a number of pivots for the binning matrix.
Some summary statistics (cross-$C_\ell$'s, WST, and WPH, in this work) can be thought of as consisting of multiple subclasses of statistics that depend exclusively on scale.
Taking WST as an example, $S_1$ and $S_2$ coefficients cannot be naturally compared based on scale.
Additionally, within the $S_2$ coefficients, statistics with different opening angles cannot be naturally ordered based on scale.
Therefore, for WST, we split the coefficients into $5$ subclasses that have only scale dependence ($S_1$ and the four different opening angle components of $S_2$). 
These statistics are ordered individually by scale and then appended to form our full ordered summary statistic.
We fix the $x$ coordinates of certain pivots to the transitions between subclasses.
This effectively introduces parameters which, in combination with the total number of bins, exclusively control how many bins are allocated to each subclass.

Pivots are therefore defined by $2$ free parameters if their $x$ coordinate is not specified and $1$ free parameter otherwise.
If there are $p$ pivots in total and $p_x$ of them have specified $x$ coordinates, the pivots can collectively be parameterized by a point in a space with dimension $2(p-p_x)+p_x$.
We call this the ``pivot space".

Each point in pivot space corresponds to a binning of the summary statistics which can be used to create a unique Fisher matrix.
To retain as much information as possible, we attempt to maximize $\det(F)$.
For the sake of efficiency, this optimization uses the Fisher matrix from Eq.\ \ref{eq:Fish} instead of the combined Fisher matrix.
The relatively low number of free parameters ($\mathcal{O}(\sim10)$ in this work) reduces the probability of overfitting compared to a general linear compression with $N\times M$ free parameters.
An outline of our optimization algorithm is given in Appendix \ref{app:bin}.

When only using one pivot, the pivot space is two-dimensional, and $\det(F)$ can be plotted as a heatmap.
An example of this for \textit{Planck} $\kappa_\text{CMB}$ $C_\ell$'s is shown in Figure \ref{fig:heatmap}.
This plot shows that $\det(F)$ has a genuine signal as a function over pivot space, but that there is also numerical noise that comes from having a finite number of simulations.
Since we maximize $\det(F)$, we could end up fitting to a local maximum in the noise, which would lead to us overestimating the strength of constraints.

To avoid overfitting, we randomly split our fiducial and derivative simulations in half, learn pivot positions on one half, and then generate contours from the other half. While we use the Fisher matrix from Eq.\ \ref{eq:Fish} to learn how to bin our summary statistics, once compressed, we use the combined Fisher matrix to create contour plots.
We repeat this process for $1000$ different splits and get a covariance matrix for our cosmological parameters for each split. 
We take our final covariance matrix to be the average of these matrices.

We use $15$ bins since we find that this choice yields binned summary statistics that follow a Gaussian distribution (see Appendix \ref{app:gauss}).
With $15$ bins, it would be feasible to determine the size of each bin individually.
However, this approach would struggle to generalize to future works if more bins were used.
Determining the optimal placements of bin boundaries for $N$ bins consists of maximizing $\det(F)$ over an $N-1$ dimensional space.
When using our pivot approach, the dimension of the pivot space, $\det(F)$, is maximized over and is completely decoupled from $N$.
This allows one to control the number of free parameters. 

In general, we note that compressing summary statistics is a non-trivial problem, and a variety of methods have been proposed in the literature \citep{Park2025}.  Many of these techniques, however, require training on simulations at varying cosmologies, which is not feasible in our setting, where only a fiducial covariance matrix and its derivatives are available. 
Other alternatives in this regime include Principal Component Analysis (PCA), finding optimal subsets of observables \citep{Szewciw_2022}, and the widely used MOPED algorithm \citep{Heavens2000}. 
In principle, MOPED would be the most principled choice, as it is designed to maximize the Fisher information.  In practice, however, its performance is limited in our case: it relies on stable estimates of derivatives and on inverting a high-dimensional covariance matrix, both of which can be noisy when only a modest number of simulations are available.  This issue is exacerbated by the need to split the simulations into independent subsets for training and validation to avoid overfitting, which further reduces the effective number available. Although our learned-binning scheme also maximizes Fisher information, the compression matrix is constrained to a simple, nearly diagonal form that depends on only a handful of pivot parameters. This structure makes it substantially less sensitive to noise than the dense compression vectors produced by MOPED, while preserving interpretability. 
Using cross-validation further mitigates overfitting to fluctuations in the Fisher matrix, yielding a robust and practical method. 

\section{Results}
\label{sec:results}

We now turn to forecasted constraints on $\Omega_m$ and $\sigma_8$ after binning our summary statistics.
We first compare auto $C_\ell$'s and WST applied to $\kappa_\text{CMB}$ maps in Section \ref{CMBL_constraints}.
Next, we look at cross-$C_\ell$'s and WPH applied to $\kappa_\text{CMB}\times\kappa_\text{WL}$ maps in Section \ref{WL_Results}.
We note that results for WST and WPH in the next section are not combined with those from the power spectrum.

\subsection{CMB Lensing}\label{CMBL_constraints}

\begin{figure}
  \includegraphics[width=0.45\textwidth]{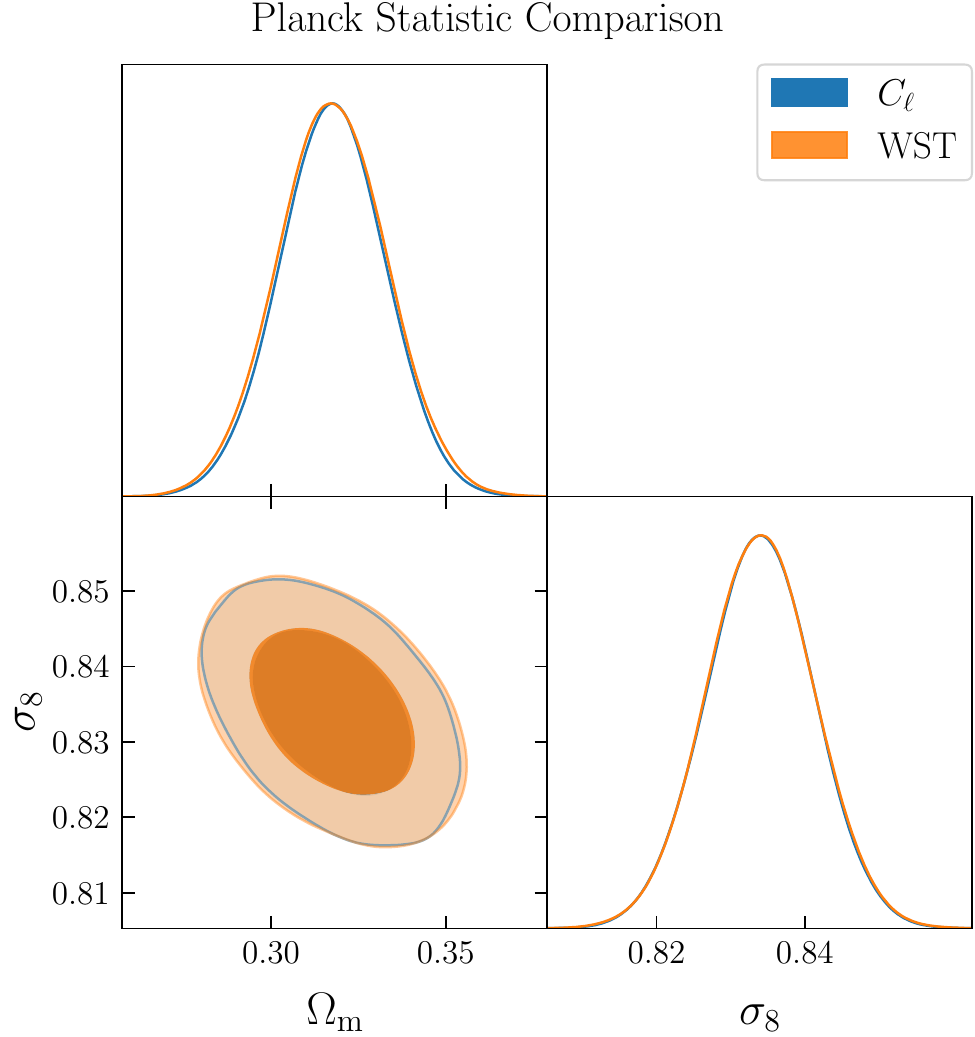}
  \caption{We show $1\sigma$ and $2\sigma$ constraints on $\Omega_m$ and $\sigma_8$ from \textit{Planck} $\kappa_\text{CMB}$, using the $C_\ell$'s (blue) and the WST coefficients (orange). These constraints are obtained using our learned-binning approach. The constraints are the same to within $5\%$.}
  \label{fig:auto}
\end{figure}

\begin{table}
    \centering
    \begin{tabular}{|l||c|c|}
        \hline
 \multicolumn{3}{|c|}{$\kappa_\text{CMB}$ constraints: $\sigma(\Omega_m)[\times100]\,\,/\,\,\sigma(\sigma_8)[\times 100]$} \\
 \hline \hline\hline\textbf{Survey} & $C_\ell$'s & WST\\ 
        \hline \hline
        
        ACT & 1.43 / 0.72 & 1.54 / 0.77\\ \hline 

        \textit{Planck} & 1.51 / 0.73 & 1.56 / 0.73 \\ \hline

        SPT & 0.90 / 0.59 & 1.00 / 0.66 \\ \hline

        SO & 0.52 / 0.35 & 0.59 / 0.40 \\ \hline
    \end{tabular}
    \caption{1$\sigma$ constraints on $\Omega_m$ and $\sigma_8$ for $C_\ell$'s and WST of $\kappa_\text{CMB}$ maps for all surveys considered in this work.}
    \label{tab:err_table}
\end{table}

For CMB lensing convergence maps, we show results for the angular power spectrum and also for WST, as our summary statistics.

For WST, we do not use dyadic sequencing; instead, we have $j$ from Eq.\ \ref{eq:Morlet} take $50$ different values from $0$ to $6$. 
This allows greater flexibility in the learned binning by increasing the size of the initial data vector.
For angular spacing, we pick $L=4$. 
With these choices, there are 5150 WST coefficients.
For $C_\ell$'s, our resolution limits the maximum $\ell$ to 3071, so after discarding the dipole, we have 3070 $C_\ell$ values. 

For the learned binning, ordering the original summary statistic is important. 
Entries that are near each other will likely be binned together, so we want these entries to be highly correlated to minimize information loss.
For $C_\ell$'s, this is done by ordering based on $\ell$.
$S_1$ coefficients are likewise ordered based on $j$.
For $S_2$ coefficients, we have two scales and an opening angle, making the ordering scheme more arbitrary.
In this work, we divide the $S_2$ coefficients into $4$ different subclasses based on the opening angle.
In each subclass, we order based on $j_2$.
When multiple coefficients have the same opening angle and $j_2$ value, we order based on $j_1$. 
Across surveys, we find that when looking at pivot positions, the $S_1$ coefficients are the most heavily weighted.

For the $C_\ell$'s, we find that allowing one pivot can slightly improve the constraints, but allowing two yields little additional gain.
Therefore, for $C_\ell$'s we search for one pivot. For WST, we lock four pivots to the transitions from $S_1$ to $S_2$ and between the different opening angles of $S_2$. 
Since one dimension of the pivot is locked, this has four degrees of freedom.
We find improvements by adding one additional arbitrary pivot that can be placed anywhere, resulting in a $ 6$-dimensional search space for the pivots, with no significant gains beyond this. 
Across surveys, we find that the first $\sim 1000$ $C_\ell$'s are consistently the most heavily weighted.

After finding the pivots, we use the combined Fisher forecast to get constraints on the cosmological parameters $\Omega_m$ and $\sigma_8$.
For all of our surveys, we find that WST and $C_\ell$'s give the same constraints on both $\Omega_m$ and $\sigma_8$ to within $15\%$, showing that there is not much non-Gaussian information to extract from $\kappa_\text{CMB}$.
\textit{Planck} constraints are shown in Figure \ref{fig:auto}, and constraints for all surveys are shown in Appendix \ref{app:all_res}. 
Additionally, all 1$\sigma$ constraints can be found in Table \ref{tab:err_table}.

When we look at the covariance of the two binned statistics, we find the expected positive correlation between $C_\ell$'s and $S_1$ coefficients.
Additionally, the covariance matrix of the WST coefficients has significant off-diagonal entries, primarily between the $S_2$ coefficients.
This highlights that while our learned binning algorithm can be useful in diagonalizing the covariance matrix, it is only possible if correlations are originally among entries close enough together to be in the same bin.
This corresponds to an initial non-binned covariance matrix with significant entries only near the diagonal.

\subsection{CMB Lensing X Weak Lensing}\label{WL_Results}

\begin{figure}
  \includegraphics[width=0.45\textwidth]{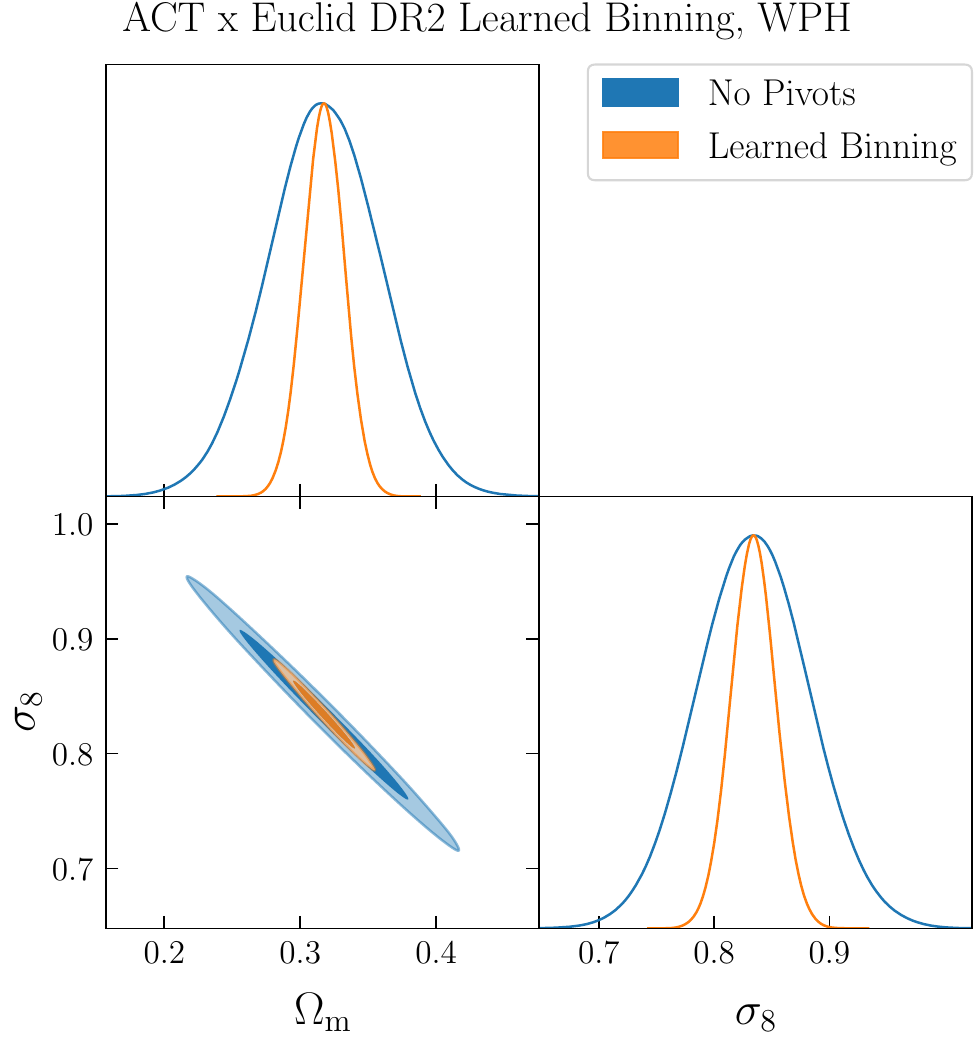}
  \caption{Constraints for ACT $\times$ \textit{Euclid} DR2 ($\kappa_\text{CMB}\times\kappa_\text{WL}$) from WPH. The learned binning (orange) outperforms the initial binning with no pivots (blue). Constraints are improved by factors of $2.7$ for $\Omega_m$ and $2.5$ for $\sigma_8$. \label{fig:learn}}
\end{figure}

\begin{figure}
  \includegraphics[width=0.45\textwidth]{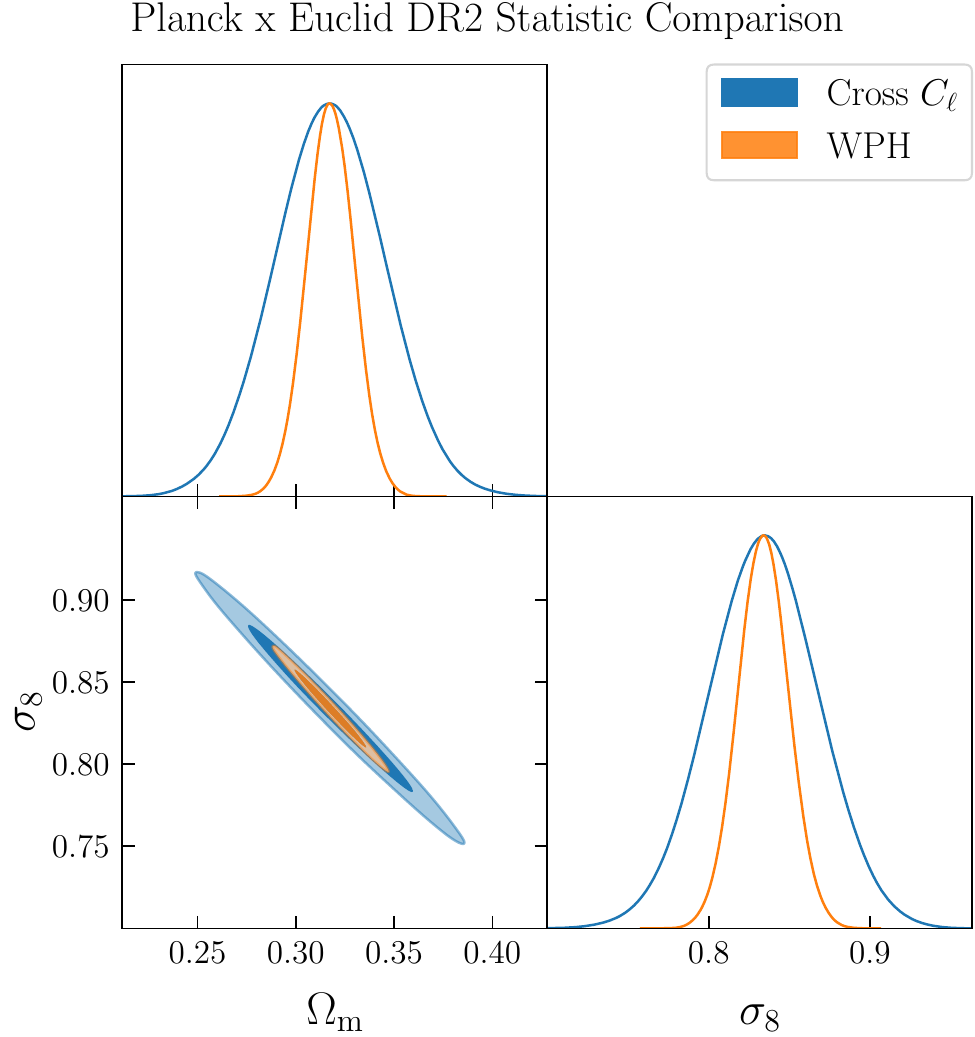}
  \caption{Constraints for \textit{Planck} $\times$ \textit{Euclid} DR2 ($\kappa_\text{CMB}\times\kappa_\text{WL}$) from cross-$C_\ell$'s and WPH using our learned-binning approach. The $\Omega_m$ and $\sigma_8$ constraints are improved by factors of $2.3$ and $2.2$, respectively, when going from cross-$C_\ell$'s to WPH.\label{fig:cross}}
\end{figure}

\begin{table}
    \centering
    \begin{tabular}{|l||c|c|}
        \hline
 \multicolumn{3}{|c|}{$\kappa_\text{CMB}\times\kappa_\text{WL}$ constraints: $\sigma(\Omega_m)[\times100]\,\,/\,\,\sigma(\sigma_8)[\times 100]$} \\
 \hline \hline\hline\textbf{Survey} & Cross-$C_\ell$'s & WPH\\ 
        \hline \hline
        
        ACT $\times$ \textit{Euclid} DR2 & 4.09 / 4.97 & 1.52 / 1.96\\ \hline 

        \textit{Planck} $\times$ \textit{Euclid} DR2 & 2.79 / 3.39 & 1.20 / 1.56 \\ \hline

        SPT $\times$ \textit{Euclid} DR2 & 2.57 / 3.29 & 0.76 / 0.95 \\ \hline

        SO $\times$ \textit{Euclid} DR2 & 2.67 / 3.39 & 0.86 / 1.10 \\ \hline
    \end{tabular}
    \caption{1$\sigma$ constraints on $\Omega_m$ and $\sigma_8$ for cross-$C_\ell$'s and WPH of $\kappa_\text{CMB}\times\kappa_\text{WL}$ maps for all surveys.}
    \label{tab:cross_err_table}
\end{table}

\begin{table}[ht]
    \centering
    \begin{tabular}{|l||c|c|}
        \hline
        \multicolumn{3}{|c|}{%
            \shortstack[c]{%
                $\kappa_\text{CMB}\times\kappa_\text{WL}$ no learned pivots constraints:\\
                $\sigma(\Omega_m)[\times100]\,/\,\sigma(\sigma_8)[\times100]$%
            }%
        } \\
        \hline\hline
        \textbf{Survey} & Cross-$C_\ell$'s & WPH \\ 
        \hline\hline
        ACT $\times$ \textit{Euclid} DR2 & 3.88 / 4.86 & 4.07 / 4.84 \\ \hline
        \textit{Planck} $\times$ \textit{Euclid} DR2 & 2.89 / 3.51 & 2.80 / 3.30 \\ \hline
        SPT $\times$ \textit{Euclid} DR2 & 2.41 / 3.08 & 1.68 / 1.96 \\ \hline
        SO $\times$ \textit{Euclid} DR2 & 2.56 / 3.24 & 1.80 / 2.16 \\ \hline
    \end{tabular}
    \caption{1 $\sigma$ constraints on $\Omega_m$ and $\sigma_8$ for cross-$C_\ell$'s and WPH of $\kappa_\text{CMB}\times\kappa_\text{WL}$ maps for all surveys. For these results, no pivots have been learned before making the constraints.}
    \label{tab:orig_err_table}
\end{table}

For the cross-correlation of the convergence of WL and CMBL fields ($\kappa_\text{CMB}\times\kappa_\text{WL}$), we use cross-$C_\ell$'s and WPH as our summary statistics.
For WPH we use $J=7$ and $L=4$, which leads to 392 coefficients. 
We use dyadic sequencing for WPH (unlike WST) to reduce runtime. Generating WPH coefficients for one map and one survey takes about $10$ minutes in our implementation.

For cross-$C_\ell$'s, we have two spectra coming from our two tomographic bins for WL, leading to 6140 $C_\ell$'s.
We order each power spectrum based on $\ell$ and then append them.
We lock a pivot at the transition between tomographic bins and add two more arbitrary pivots so that each tomographic bin can get one pivot like the auto $C_\ell$'s of the previous section.
This leads to a $ 5$-dimensional search space for the pivots, and we find no significant improvements beyond this.
Across surveys, we find that the first $\sim 500$ $C_\ell$'s from the higher redshift tomographic bin are the most heavily weighted.

For WPH, for each tomographic bin we have $S_{00}$, $S_{01}$, $S_{10}$, $S_{11}$, $C_{01}$, and $C_{10}$ coefficients.
For $C_{01}$ and $C_{10}$, we also have four different opening angles that we split into different subclasses of statistics.
From this, for each of the two tomographic bins, we have 12 subclasses within the WPH summary statistic, leading to a total of 24 subclasses.
Within each subclass, we order based on $j$.
For $C_{01}$ ($C_{10}$), we order based on $j_2$ ($j_1$).
When multiple coefficients have the same opening angle and $j_2$ ($j_1)$ value, we order based on $j_1$ ($j_2$).
We lock pivots at the transitions between each subclass of the WPH statistics.
Since this configuration already has a relatively large number of degrees of freedom for binning, we do not add any additional pivots.
Across surveys, $S_{00}$ and $S_{11}$ consistently receive the largest number of bins.

As an example of the power of the learned binning, we can compare constraints with the learned binning compared to an initial binning.
As the simplest possible initial binning to compare to, we look at linear binning with no pivots. Constraints are shown in Figure \ref{fig:learn} for ACT WPH.
Constraints improve by factors of $2.7$ for $\Omega_m$ and $2.5$ for $\sigma_8$ when using a learned binning.
We verify the convergence of this method in Appendix \ref{app:conv_move}.

We find that constraints on $\Omega_m$ and $\sigma_8$ are both significantly tighter when using WPH compared to cross-$C_\ell$'s. 
Constraints on $\Omega_m$ improve by factors between $2.3$ for \textit{Planck} $\times$ \textit{Euclid} DR2 and $3.4$ for SPT $\times$ \textit{Euclid} DR2.
Improvements for $\sigma_8$ are between factors of $2.2$ for \textit{Planck} $\times$ \textit{Euclid} DR2 and $3.4$ for SPT $\times$ \textit{Euclid} DR2. 
In Figure \ref{fig:cross} we show this result for \textit{Planck} $\times$ \textit{Euclid} DR2.
Additionally, all 1$\sigma$ constraints can be found in Table \ref{tab:cross_err_table}.
Full contour results for other surveys are provided in Appendix \ref{app:all_res}. 
All 1$\sigma$ constraints when no pivots are learned are also presented in Table \ref{tab:orig_err_table}.
Cross-$C_\ell$'s see no significant changes in constraining power when using learned binning, with constraints fluctuating by under $10\%$.
For our CMB lensing auto statistics, $C_\ell$'s also see constraint changes of under $10\%$ when using learned binning.
Larger impacts are observed with WST, where constraints on $\Omega_m$ are up to $\sim1.5$ times tighter in ACT.
Results do not change appreciably for the power spectrum statistics if logarithmic binning is used instead of linear binning.

Our results show that WPH can extract information from non-Gaussianities in the lensing. 
Compared to the lack of gains found for CMB lensing alone, this is expected: weak-lensing -- CMB-lensing cross-correlations peak at significantly lower redshift ($z\sim1$) compared to CMB lensing auto-correlations ($z\sim2$–6). 
At these lower redshifts, the matter distribution is more evolved and inherently more non-Gaussian, providing WPH with additional information to extract beyond that available to power spectra alone. 
These findings are consistent with those of previous WL applications in the literature \citep[e.g.][]{Cosmo_WST, DES_WST, HSC_WST, prat2025, jeffrey2025}.

To test the convergence of our results, we repeat this entire process (including finding pivots) while lowering the total number of derivatives used.
We examine how the constraints vary with the number of derivative realizations used and find that the results have sufficiently converged.
All results are shown in Appendix \ref{app:conv_move}.

\section{Conclusion}
\label{sec:conclusion}

In this work, we study CMB lensing convergence and galaxy weak-lensing convergence data using a variety of summary statistics. 
To reduce the dimensionality of our summary statistics, we introduce a novel, learned-binning approach.
Like other data reduction methods such as MOPED \citep{Heavens2000}, our algorithm attempts to maximize Fisher information in the compressed statistic.
Since our method uses fewer parameters, it is less susceptible to noise from a limited number of simulations.
Our method also never requires inverting the original data covariance matrix, which becomes impossible if the number of simulations is less than the length of the summary statistic.
By construction, our learned binning approach is robust to overfitting and can improve constraints relative to linear binning by up to a factor of 2.7 for the surveys and observables considered in this work.

For CMB lensing convergence data, we compare the higher-order statistics of the wavelet scattering transform to the standard angular power spectrum.
We find that these summary statistics lead to similar constraints for all of the surveys considered.
When we cross our CMB lensing simulated data with \textit{Euclid} DR2 galaxy weak lensing convergence, we find that the higher-order statistic of the wavelet phase harmonics significantly outperforms the cross power spectrum.
Constraints are improved by factors of up to $3.4$.
Compared to constraints from CMB lensing convergence data alone, the cross-survey constraints exhibit stronger degeneracies.
All contour plots are presented in Appendix \ref{app:all_res}.

With WPH showing the most promise of the two higher-order statistics, further generalizations of WPH could prove useful.
One potential next step would be investigating the use of WPH statistics besides $S_{00}$, $S_{01}$, $S_{10}$, $S_{11}$, $C_{01}$, and $C_{10}$.
WPH and WST also require us to break the sky into patches, which limits the scales we can probe.
Generalizing our pipeline to spherical wavelets~\citep{2021arXiv210202828M} could circumvent this issue. Spherical wavelets have been previously used for CMB analysis \citep{McEwen:2007ni,surrao2024constrainingcosmologicalparametersneedlet}. 

This work demonstrates that, even though higher-order statistics are largely unnecessary for CMB lensing applications, they can substantially improve cosmological parameter constraints from galaxy weak-lensing--CMB-lensing cross-correlation data.

We also provide a novel method for comparing multiple summary statistics without sacrificing interpretability, which will become increasingly important as new summary statistics are introduced.

A physical interpretation of these results follows from the discussion in the previous section. WL-CMBL cross-correlations peak at lower redshifts ($z\sim1$) than CMBL convergence auto-correlations ($z\sim2$–6), where the matter distribution is more evolved and therefore more non-Gaussian. This makes non-Gaussian statistics particularly powerful in the cross-correlation case. While applications of non-Gaussian statistics to weak lensing alone are already well established \citep[e.g.][]{Cosmo_WST, DES_WST, HSC_WST, Chudaykin:2025aux, sunao2025, prat2025, jeffrey2025}, their use in WL–CMBL convergence cross-correlation data has not been explored before. One reason is that it requires forward modeling of the CMB lensing reconstruction process, including all relevant observational systematics, which is highly non-trivial. Nonetheless, the results presented here suggest that such an effort would be worthwhile.

\section{Acknowledgements}
\label{sec:Acknowledgements}

The authors would like to thank Anmol Raina for his contributions during the initial stages of this project and for insightful discussions. GV acknowledges the support of the Eric and Wendy Schmidt AI in Science Fellowship at the University of Chicago, a program of Schmidt Sciences. MG is  supported by the Kavli Institute for Cosmological Physics at the University of Chicago
through an endowment from the Kavli Foundation.

\clearpage

\onecolumngrid

\appendix
\numberwithin{figure}{section}
\numberwithin{table}{section}

\section{Constraints for All Surveys}
\label{app:all_res}

In this Appendix, we show all the parameter constraints studied in this work. All the constraints shown here are obtained using our learned-binning approach.

\begin{figure}[htbp]
  \centering
  \resizebox{\linewidth}{!}{
    \begin{tabular}{cc}
      \includegraphics[width=0.48\linewidth]{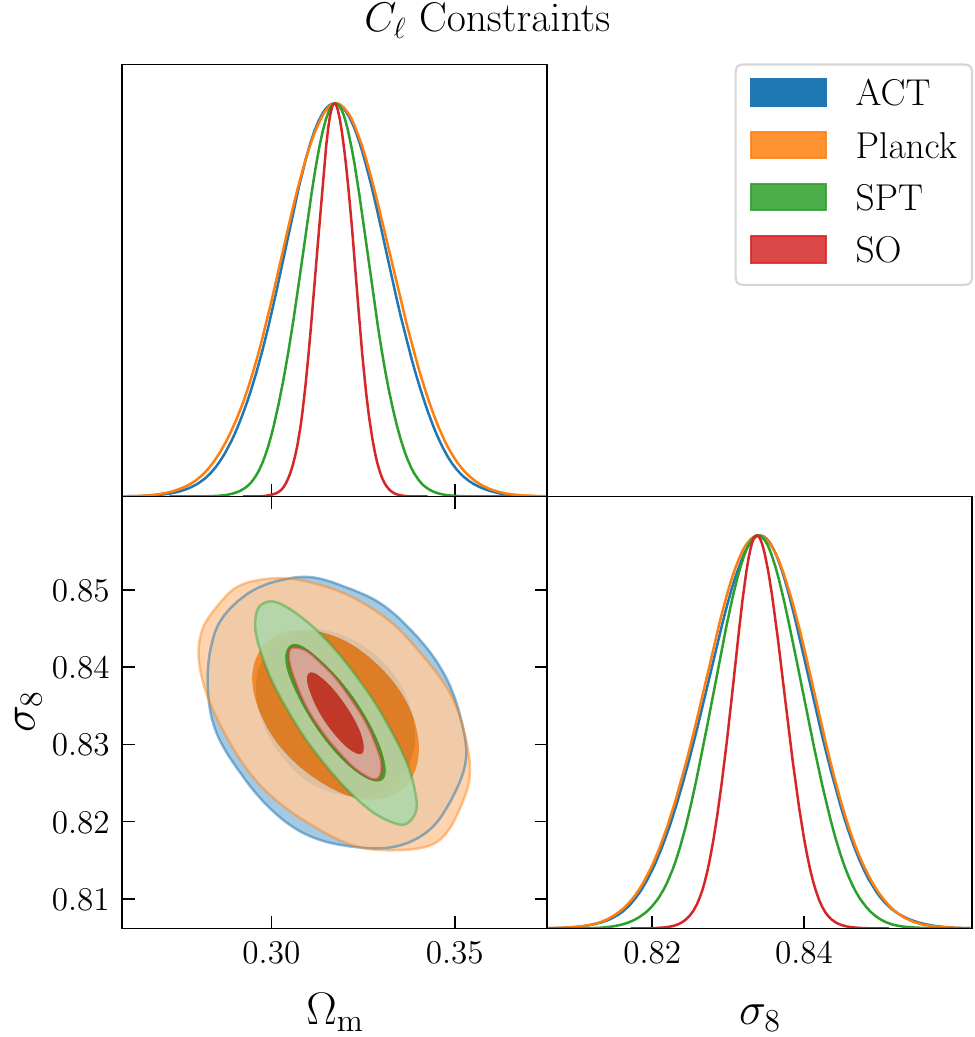} &
      \includegraphics[width=0.48\linewidth]{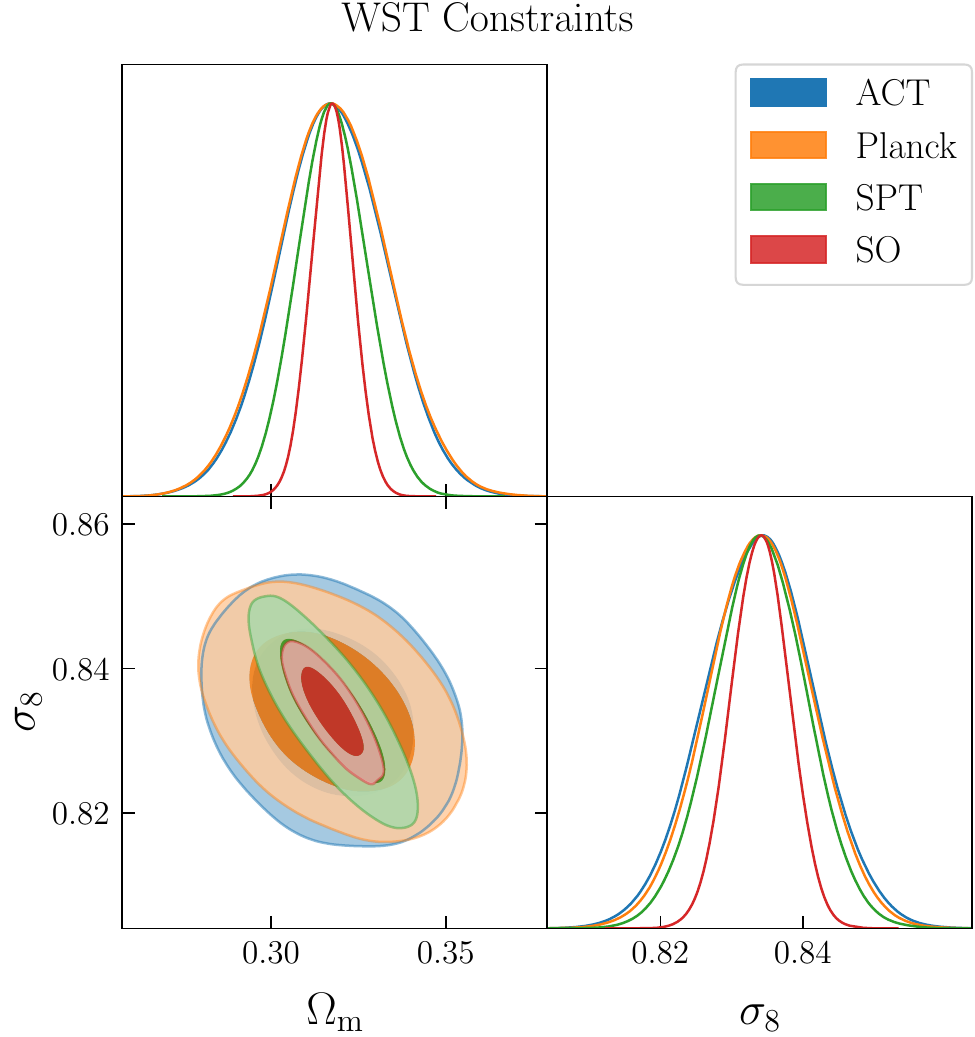} \\[1.5em]  
      \includegraphics[width=0.48\linewidth]{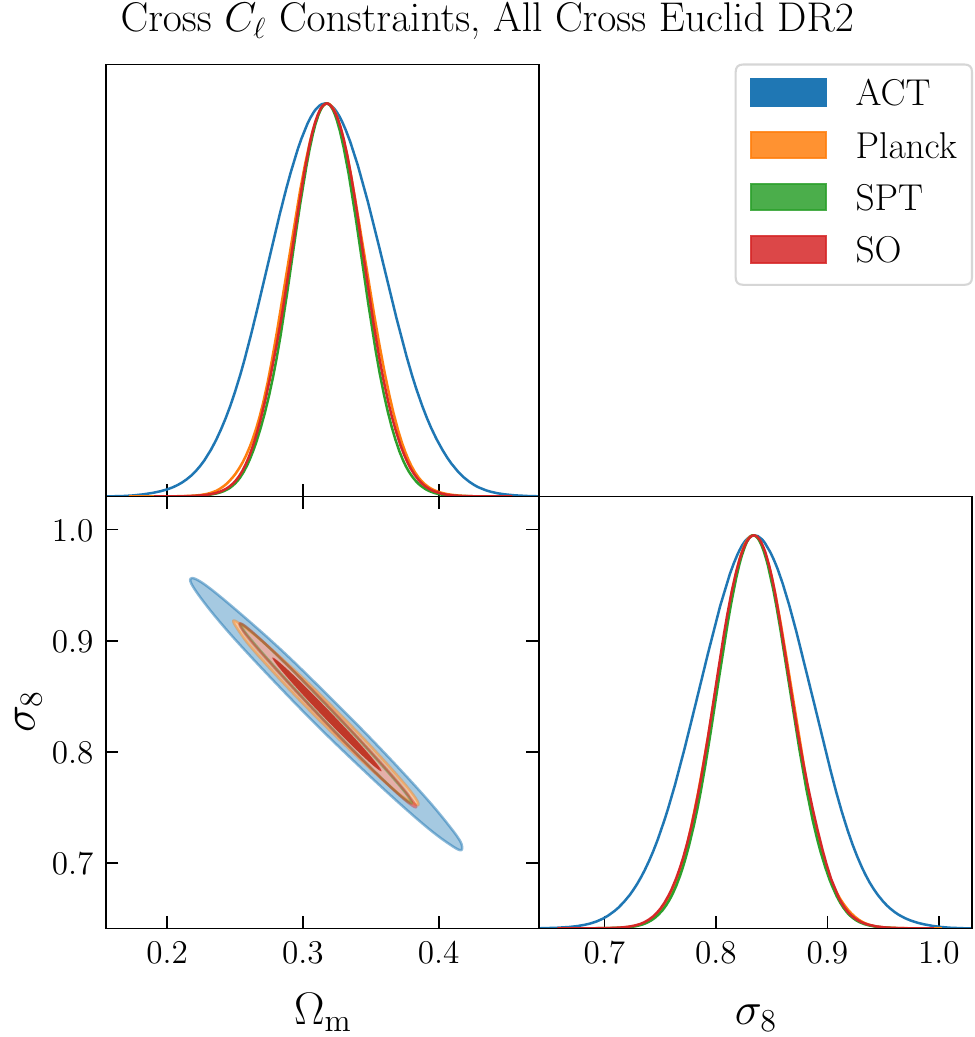} &
      \includegraphics[width=0.48\linewidth]{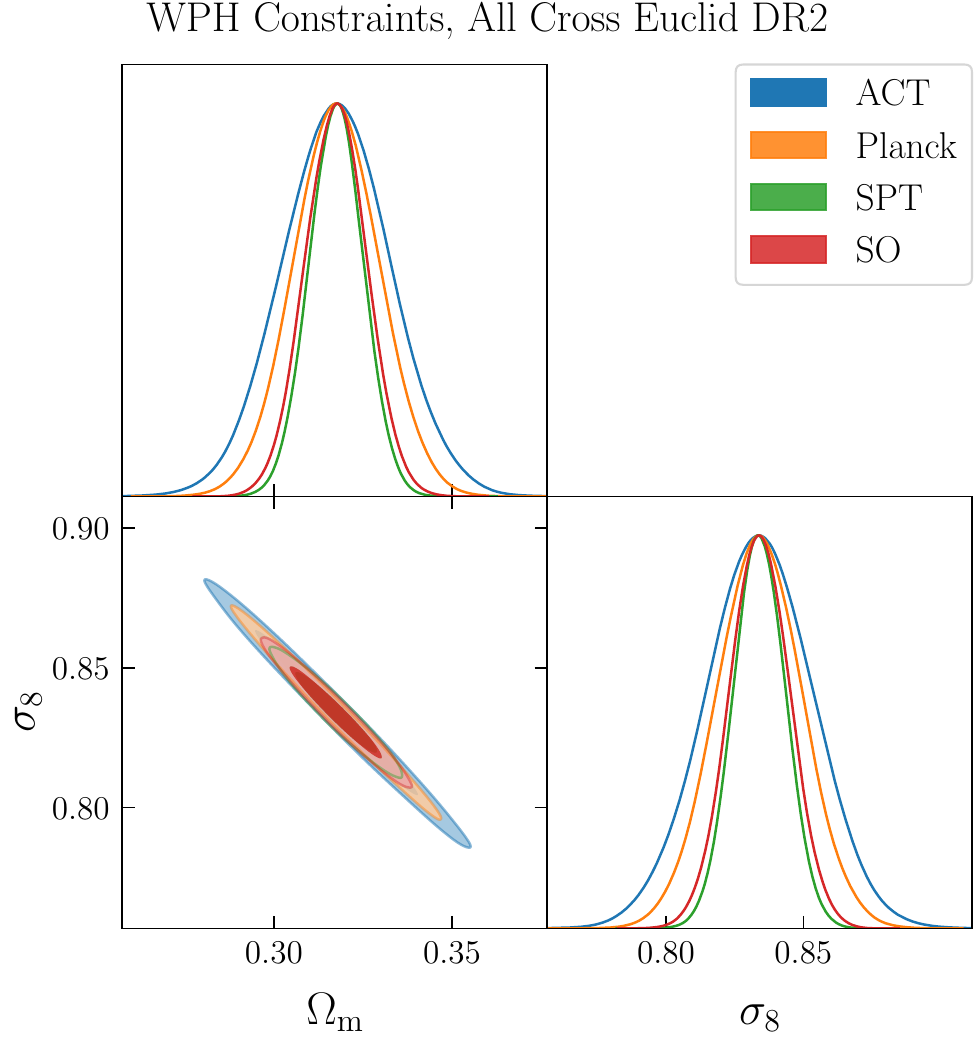}
    \end{tabular}
  }
  \caption{Constraint comparisons between surveys for all summary statistics used.
  The upper plots use $\kappa_\text{CMB}$ while the bottom plots use $\kappa_\text{CMB}\times\kappa_\text{WL}$.}
\end{figure}

\begin{figure}[htbp]
  \centering
  \resizebox{\linewidth}{!}{
    \begin{tabular}{cc}
      \includegraphics[width=0.48\linewidth]{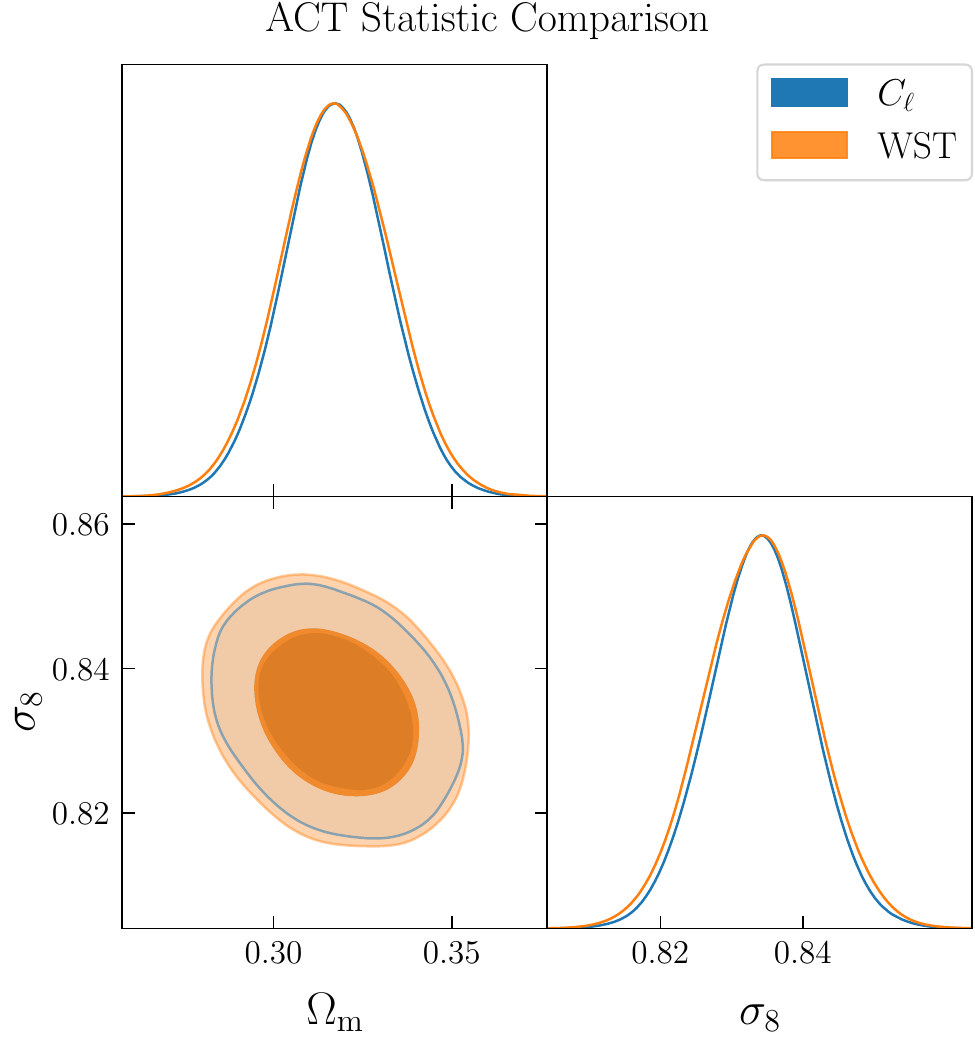} &
      \includegraphics[width=0.48\linewidth]{figures/Paper_Plots_v2/Planck.pdf} \\[1.5em]  
      \includegraphics[width=0.48\linewidth]{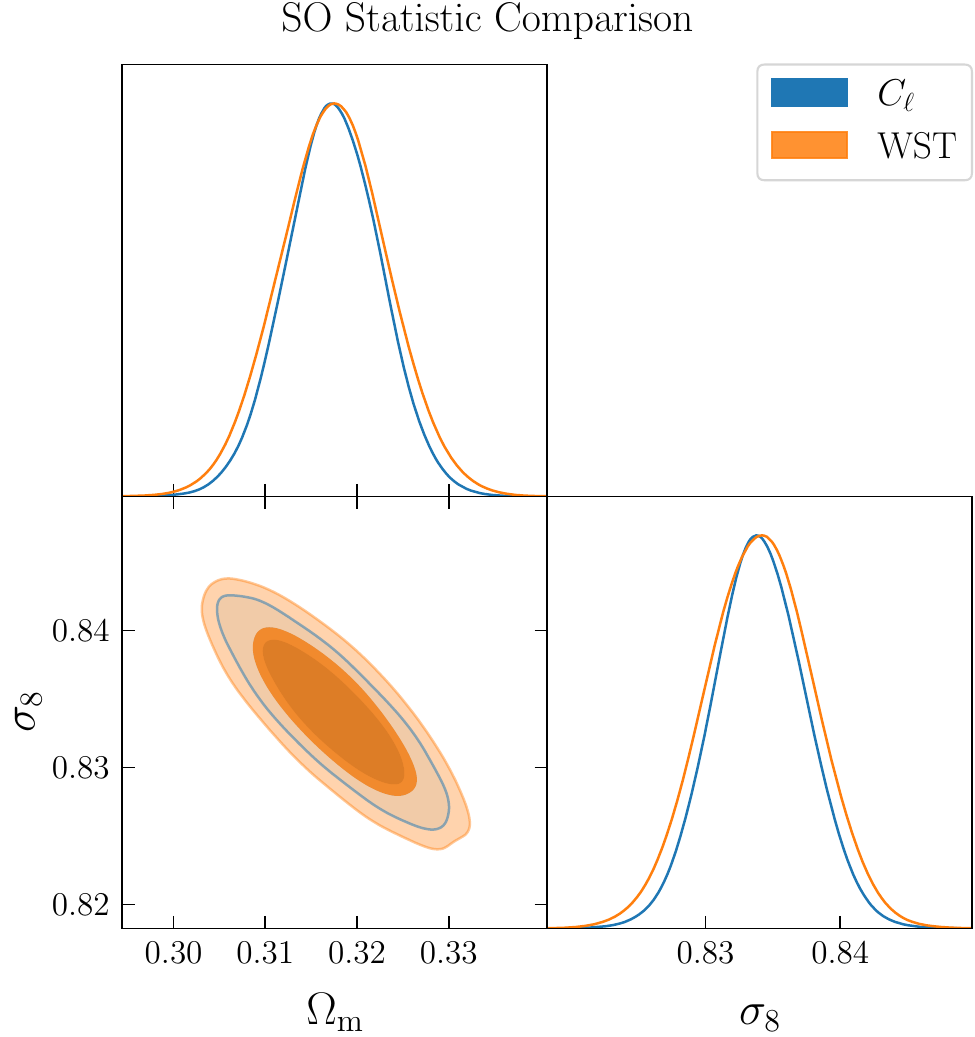} &
      \includegraphics[width=0.48\linewidth]{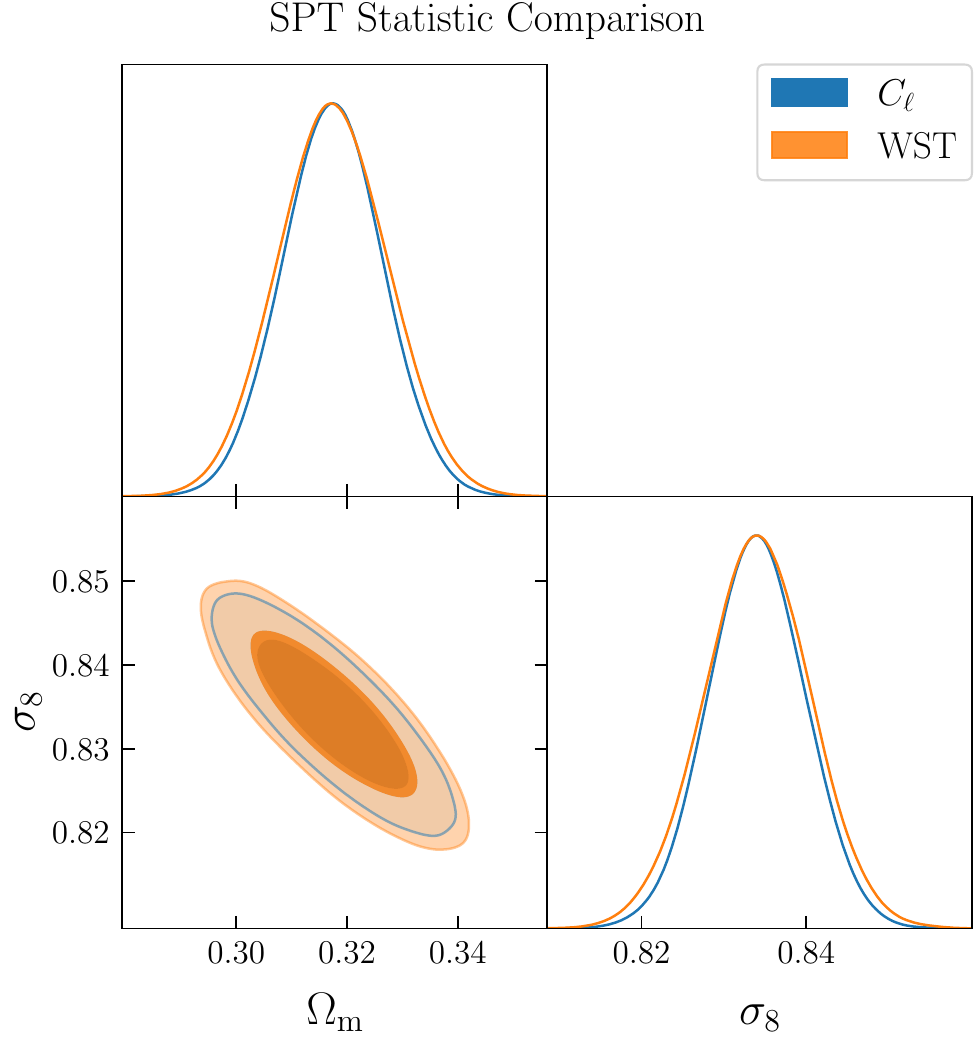}
    \end{tabular}
  }
  \caption{Constraint comparisons between $C_\ell$'s and WST for $\kappa_\text{CMB}$ maps.}
\end{figure}

\begin{figure}[htbp]
  \centering
  \resizebox{\linewidth}{!}{
    \begin{tabular}{cc}
      \includegraphics[width=0.48\linewidth]{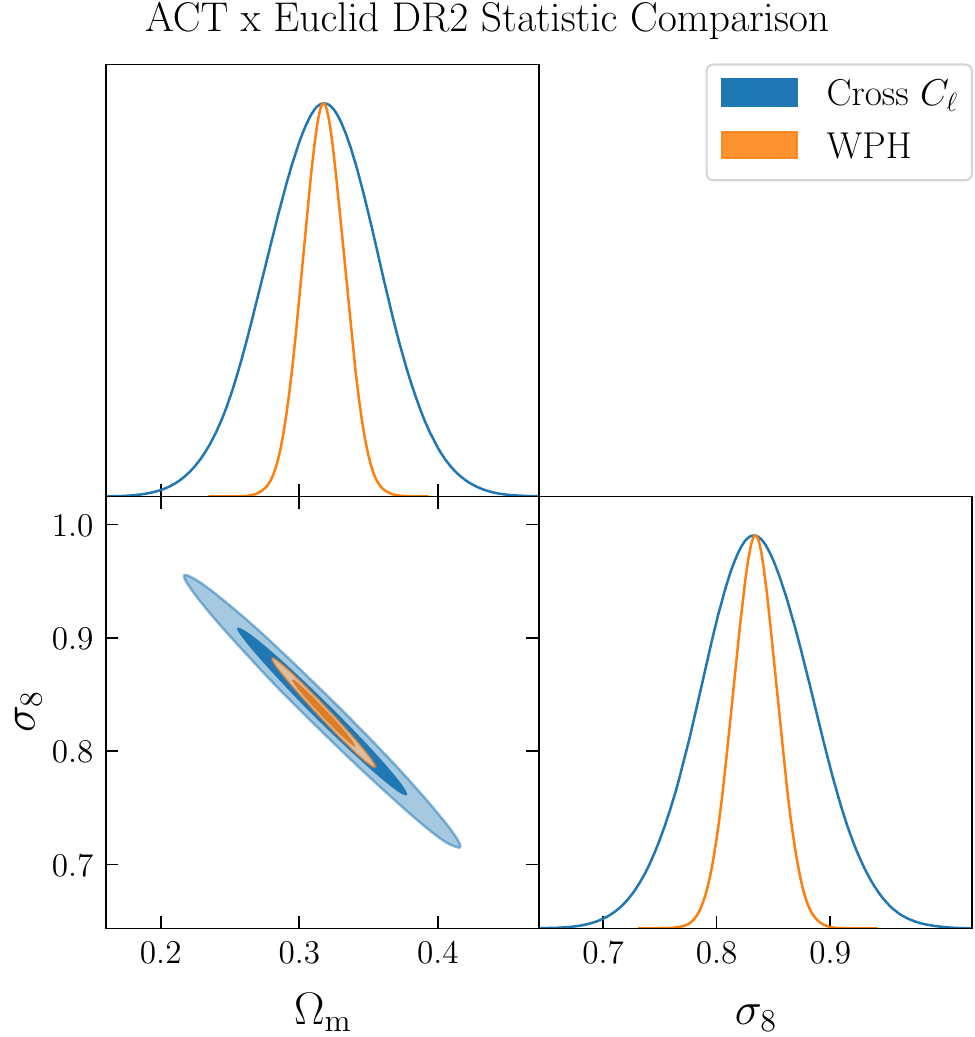} &
      \includegraphics[width=0.48\linewidth]{figures/Paper_Plots_v2/Planck_x_DR3.pdf} \\[1.5em]  
      \includegraphics[width=0.48\linewidth]{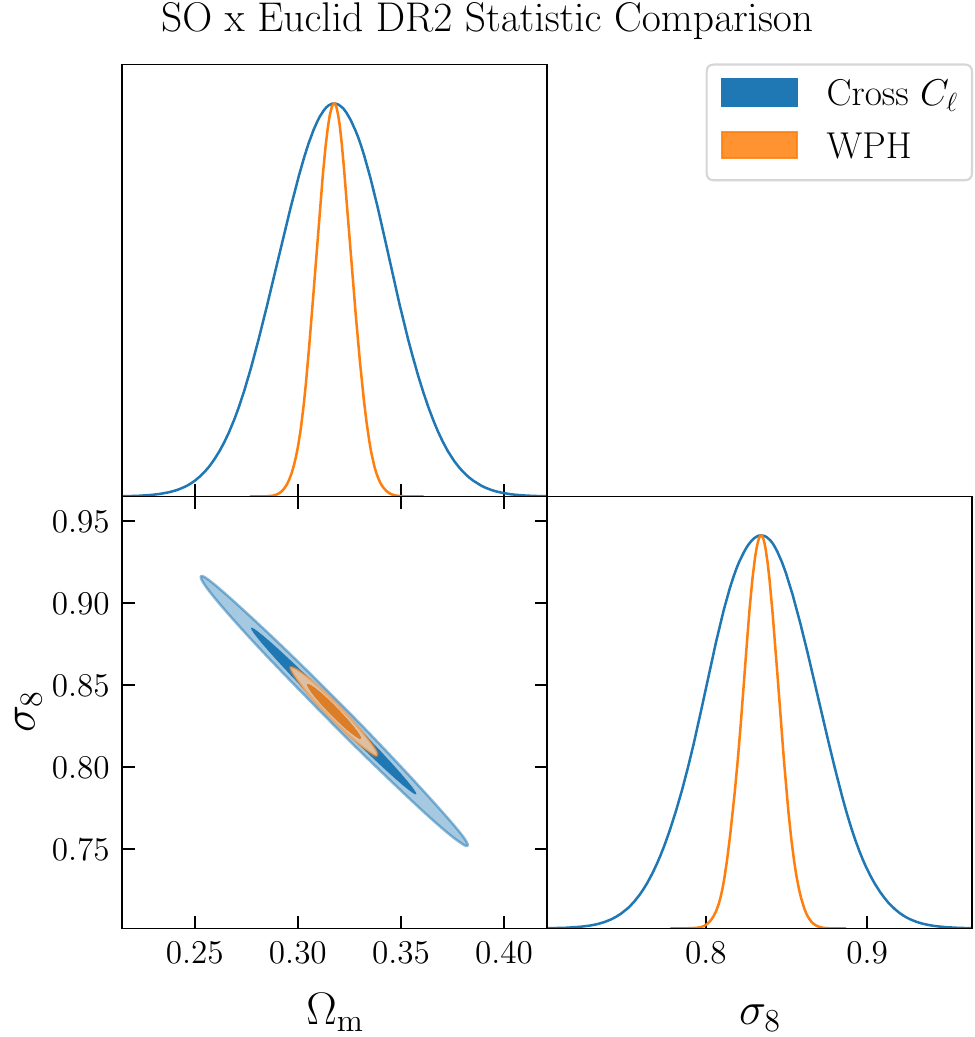} &
      \includegraphics[width=0.48\linewidth]{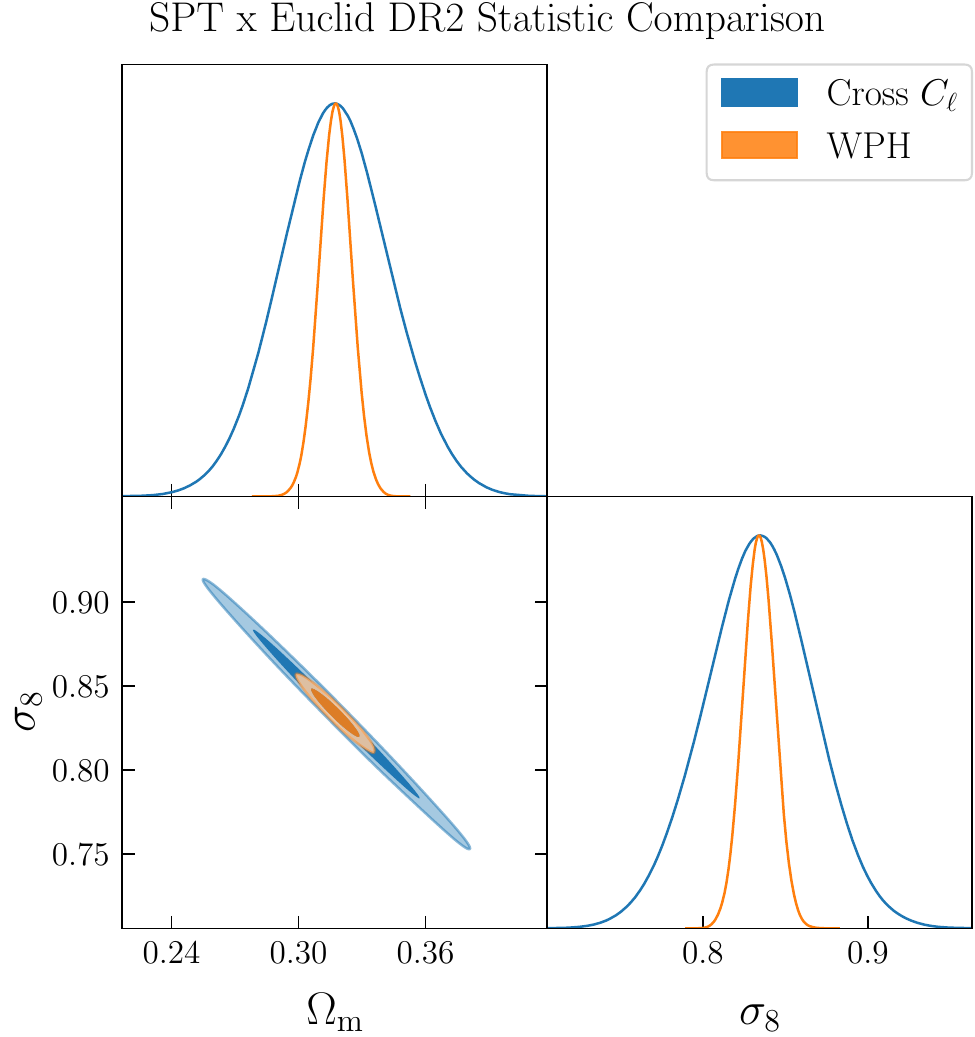}
    \end{tabular}
  }
  \caption{Constraint comparisons between cross-$C_\ell$'s and WPH for $\kappa_\text{CMB}\times\kappa_\text{WL}$.}
\end{figure}
\clearpage

\section{Contour Convergence for All Surveys}
\label{app:conv_move}
%\FloatBarrier

\begin{figure}[htbp]
  \centering
  \resizebox{\linewidth}{!}{
    \begin{tabular}{cc}
      \includegraphics[width=0.48\linewidth]{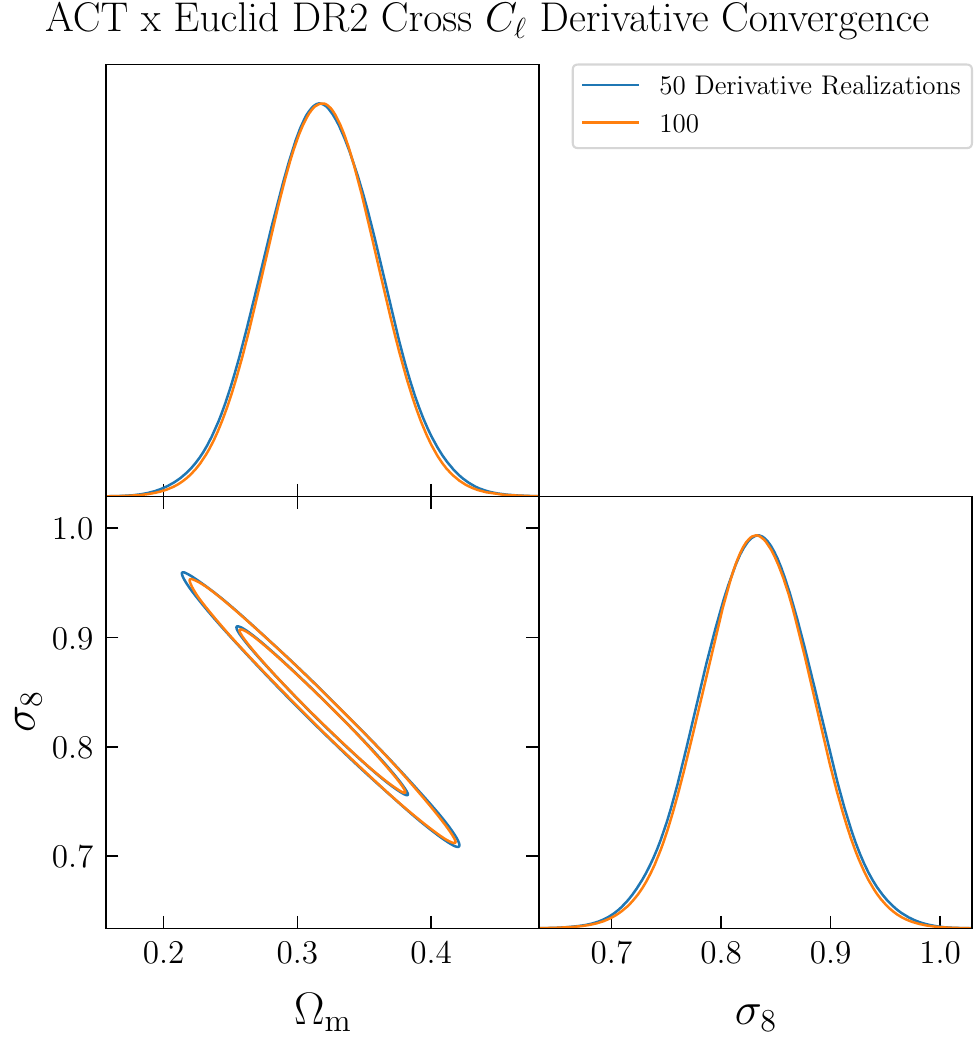} &
      \includegraphics[width=0.48\linewidth]{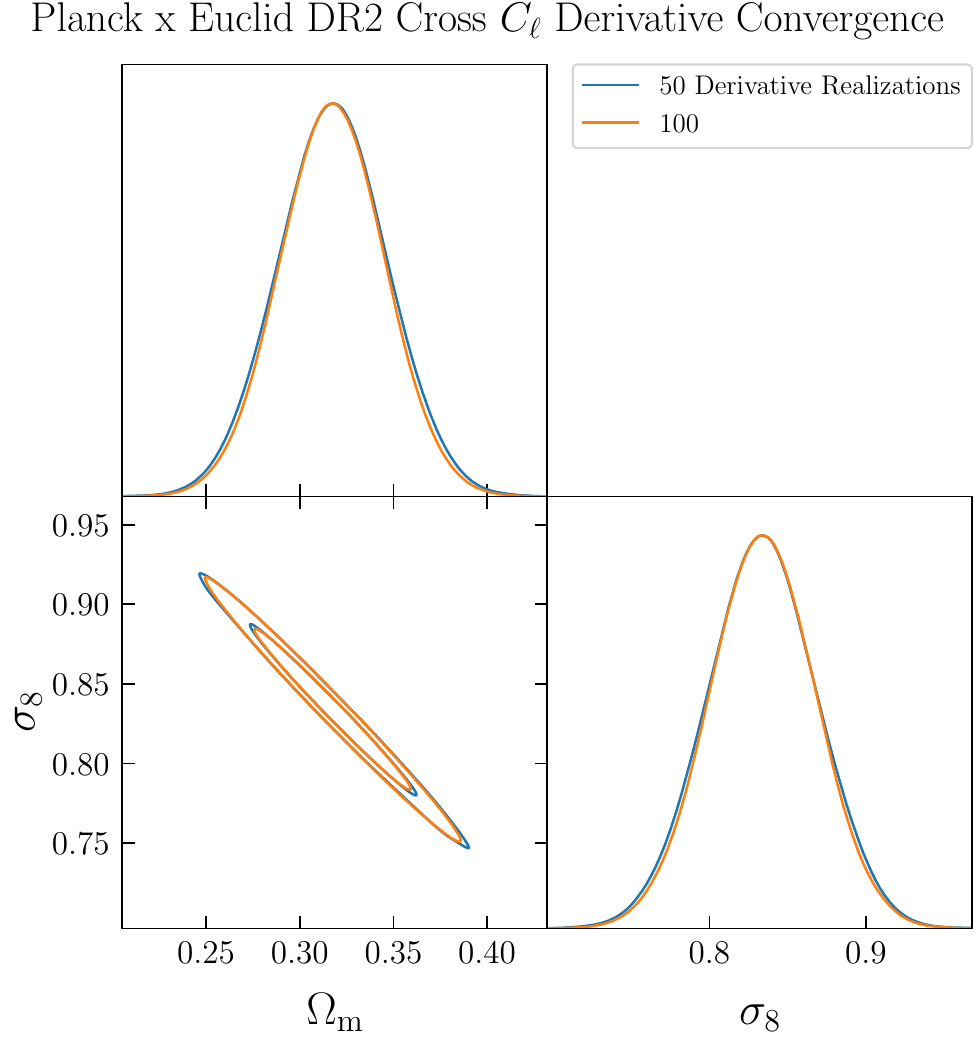} \\[1.5em]  
      \includegraphics[width=0.48\linewidth]{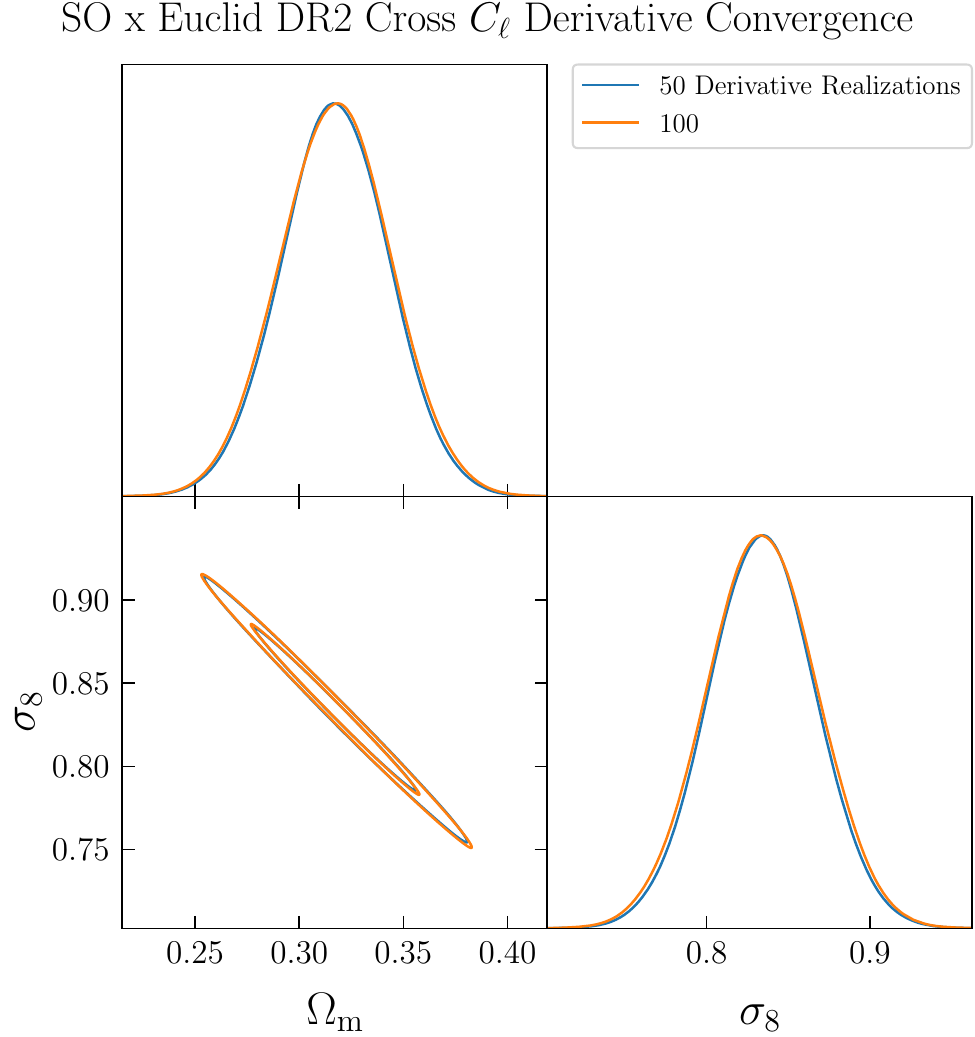} &
      \includegraphics[width=0.48\linewidth]{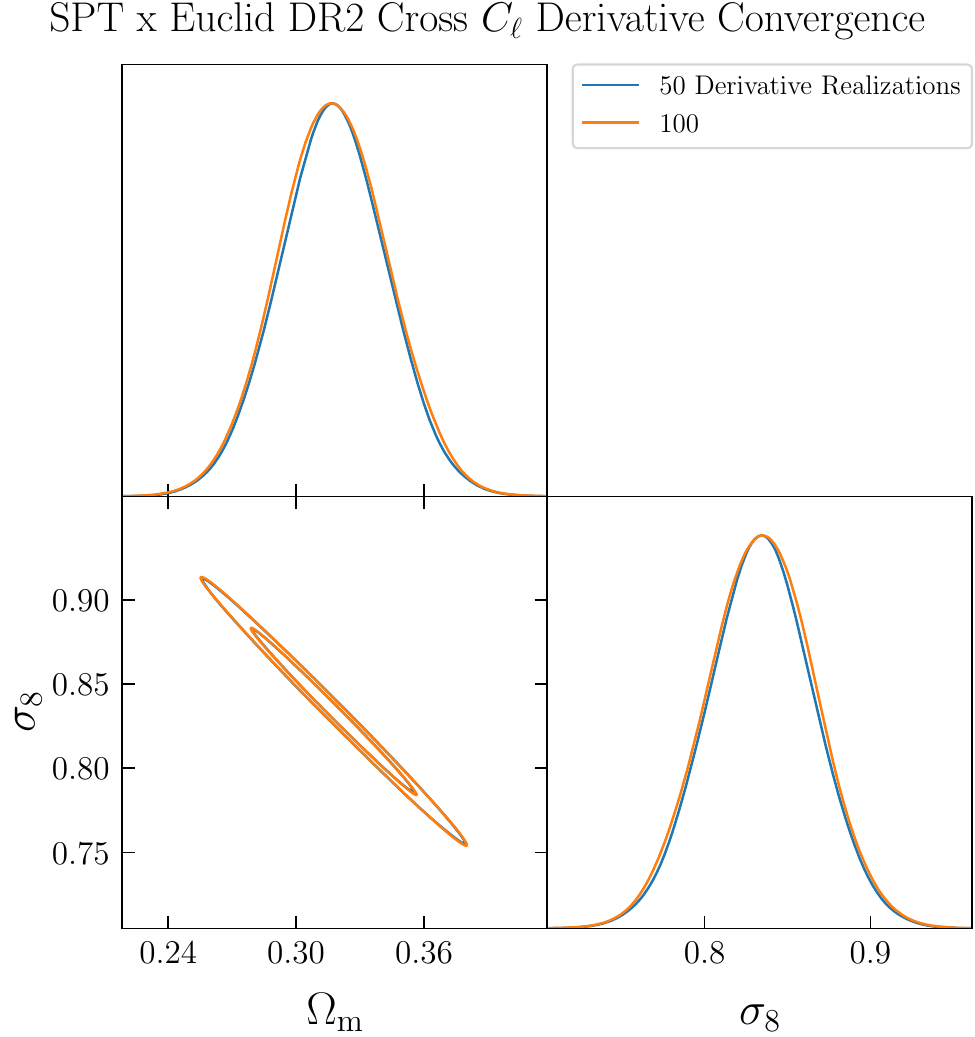}
    \end{tabular}
  }
  \caption{Contour convergence as fewer derivatives are used for cross-$C_\ell$'s applied to $\kappa_\text{CMB}\times\kappa_\text{WL}$. Shown are contours when 50\% (blue) and 100\% (orange) of the total available derivatives are used.}\label{fig:convergence_der1}
\end{figure}

\begin{figure}[htbp]
  \centering
  \resizebox{\linewidth}{!}{
    \begin{tabular}{cc}
      \includegraphics[width=0.48\linewidth]{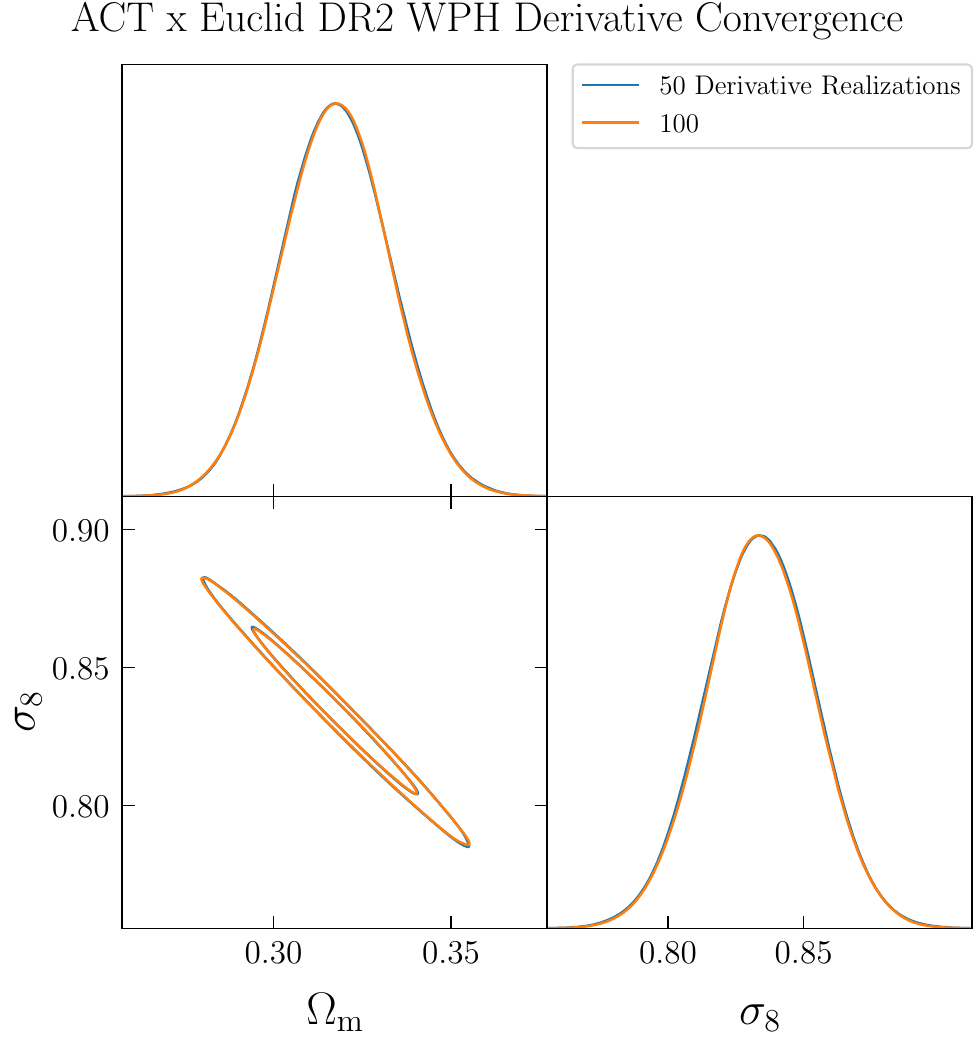} &
      \includegraphics[width=0.48\linewidth]{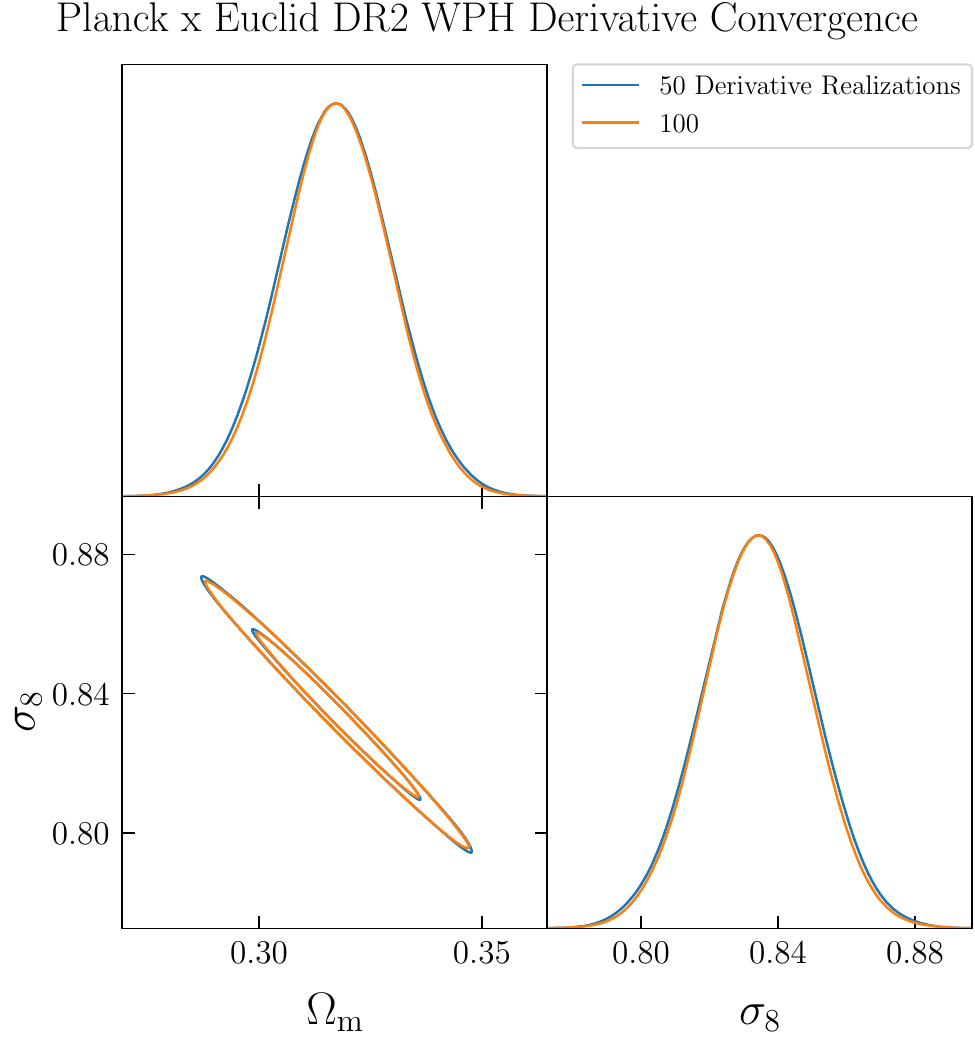} \\[1.5em]  
      \includegraphics[width=0.48\linewidth]{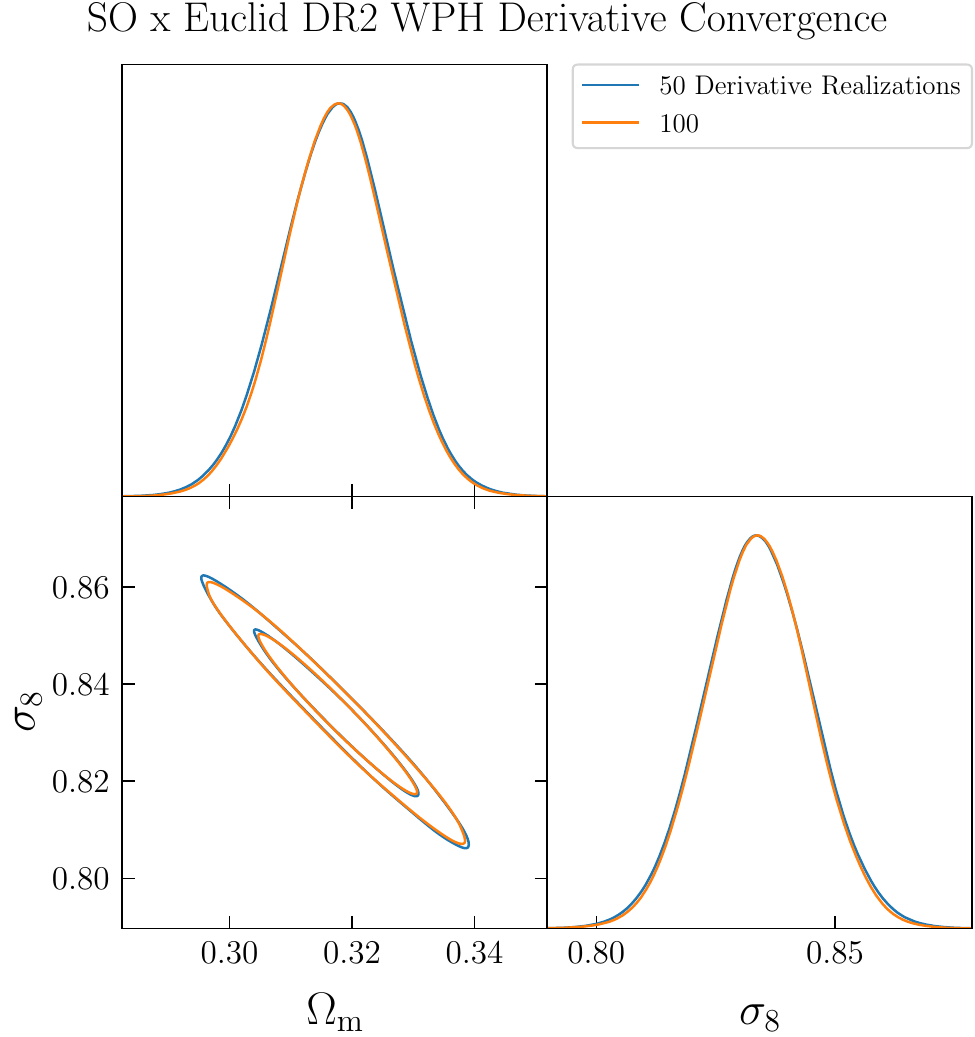} &
      \includegraphics[width=0.48\linewidth]{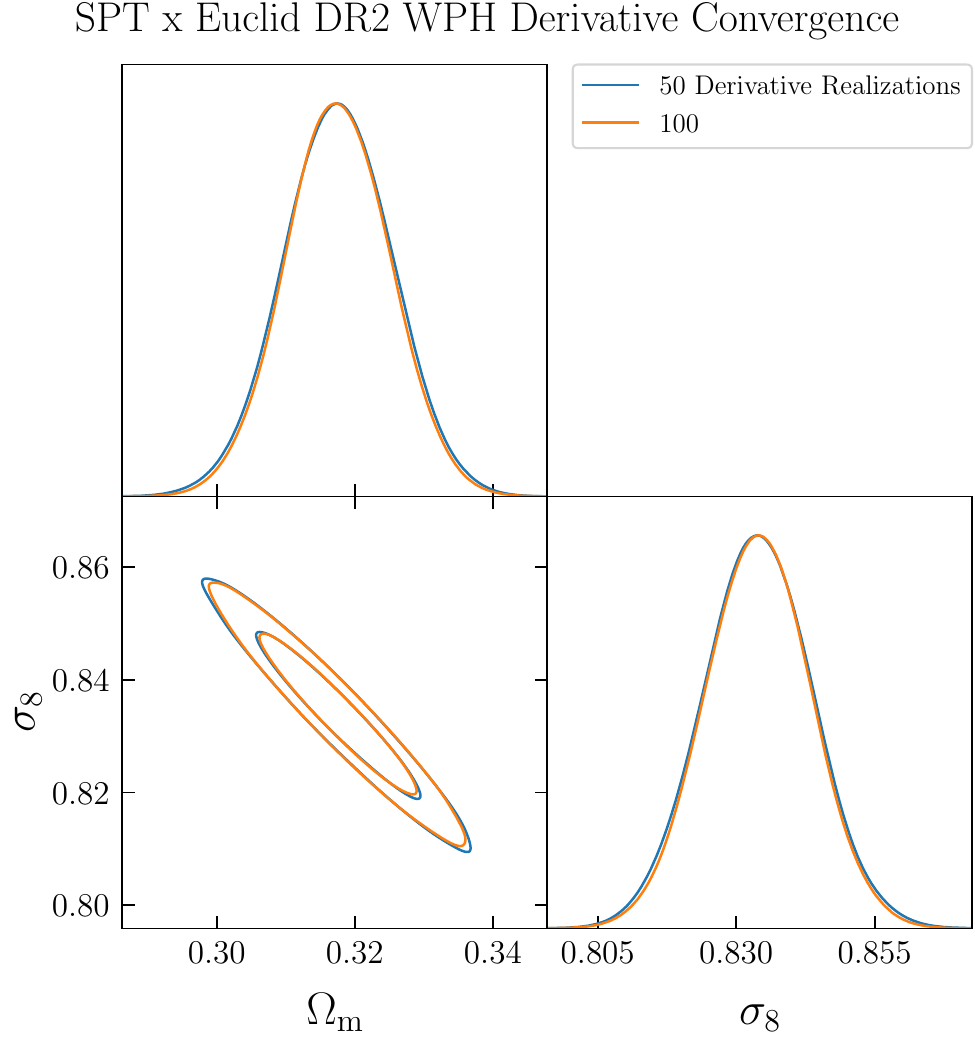}
    \end{tabular}
  }
  \caption{Contour convergence as fewer derivatives are used for WPH applied to $\kappa_\text{CMB}\times\kappa_\text{WL}$. Shown are contours when 50\% (blue) and 100\% (orange) of the total available derivatives are used.}\label{fig:convergence_der2}
\end{figure}

\begin{figure*}
    \centering
  \includegraphics[width=0.90\textwidth]{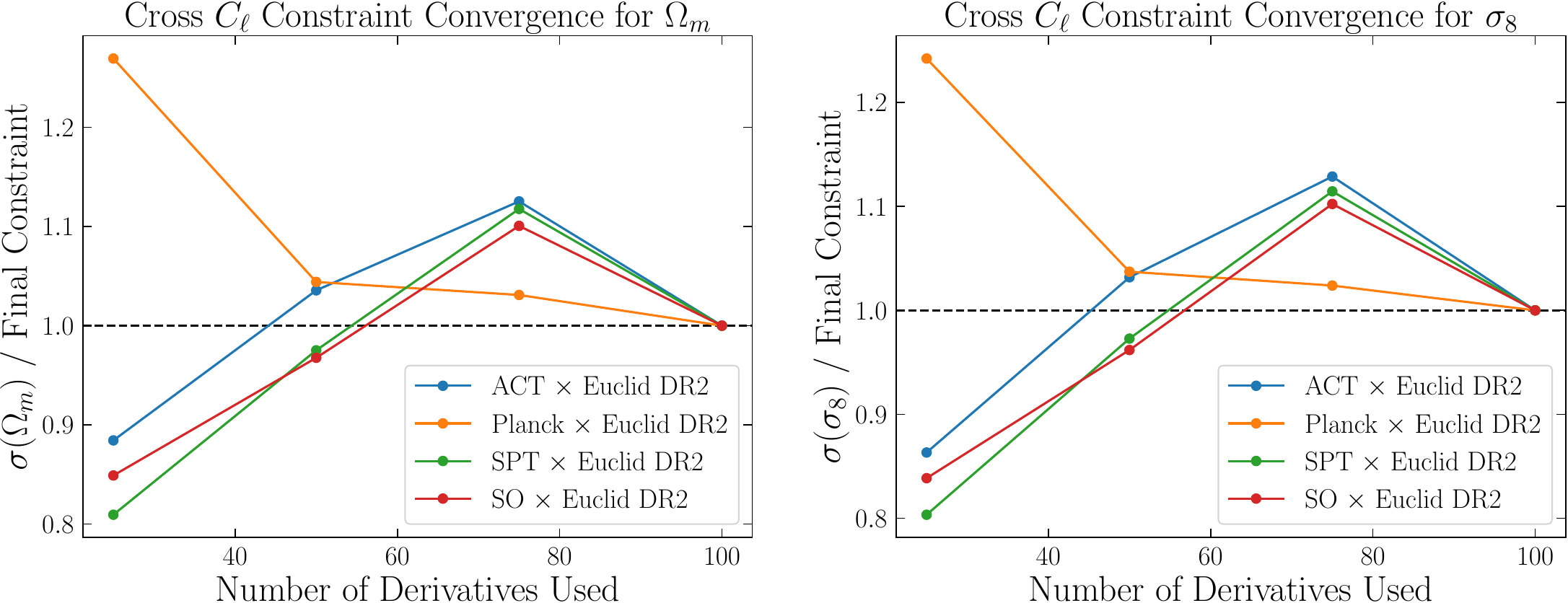}
  \caption{Convergence of $\Omega_m$ (left) and $\sigma_8$ (right) constraints as a function of the number of derivative realizations used to evaluate cross-$C_\ell$'s from $\kappa_\text{CMB}\times\kappa_\text{WL}$. By the time 50 realizations are used, constraints have converged to within $\sim 15\%$ of their final values. The high degeneracy of the parameters makes the two plots look nearly identical.}\label{fig:cro_conv}
\end{figure*}

\begin{figure*}
    \centering
  \includegraphics[width=0.90\textwidth]{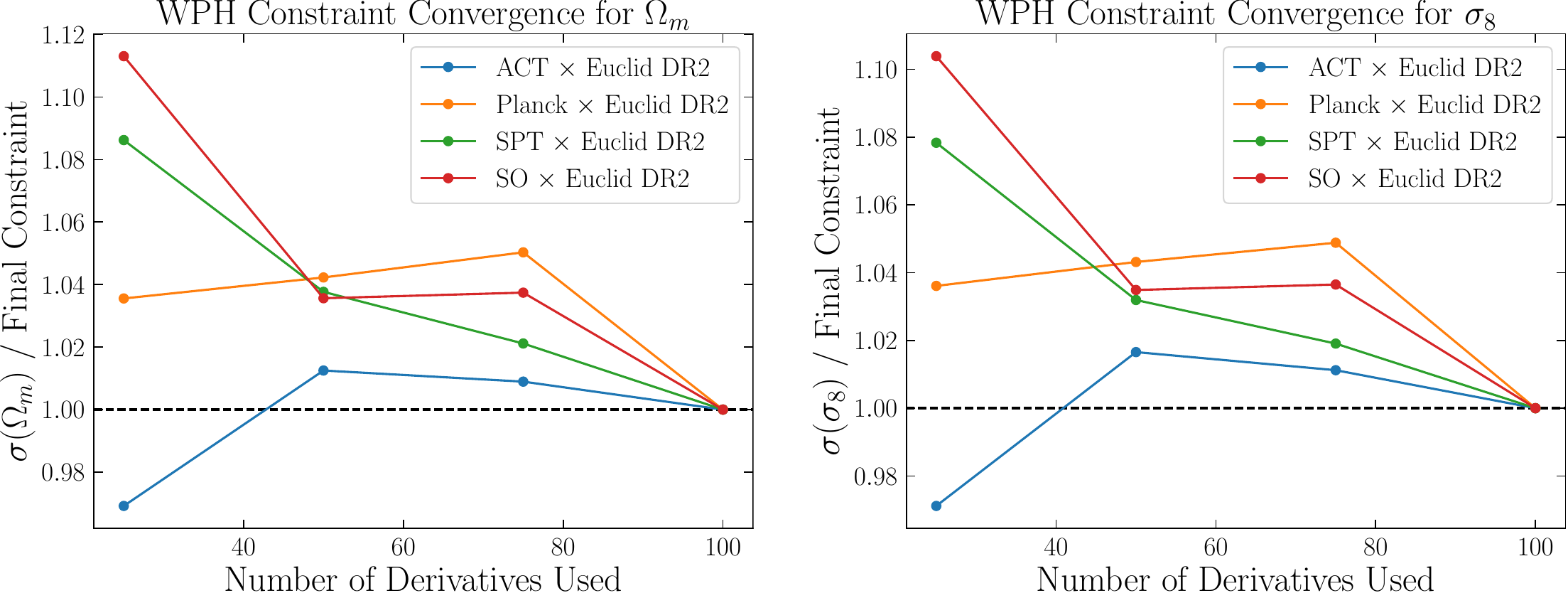}
  \caption{Same as Figure \ref{fig:cro_conv}, but using WPH as the summary statistic. By the time 50 realizations are used, constraints have converged to within $\sim 5\%$ of their final values.}\label{fig:wph_conv}
\end{figure*}

We test the convergence of our methodology by reducing the number of available derivatives and observing how the contours change. The number of realizations for generating both $\Omega_m$ and $\sigma_8$ derivatives is changed simultaneously.
For brevity, we only show the convergence contours for $\kappa_\text{CMB}\times\kappa_\text{WL}$ statistics in Figures \ref{fig:convergence_der1} and \ref{fig:convergence_der2}.
Constraints for auto $\kappa_\text{CMB}$ statistics all converge to similar levels. 
Additionally, we show how $1\sigma$ constraints on $\Omega_m$ and $\sigma_8$ converge individually for cross-$C_\ell$'s and WPH, in Figures \ref{fig:cro_conv} and \ref{fig:wph_conv}, respectively.

\clearpage

\section{Derivative Convergence}
\label{app:conv_ders}

This section focuses only on the convergence of derivatives of our summary statistics with respect to $\Omega_m$.
This is done for the sake of brevity; $\sigma_8$ derivatives converge to similar levels.
We have sets of maps for three different values of $\Omega_m$: the fiducial value and the fiducial value $\pm\Delta\Omega_m$ with $\Delta\Omega_m=0.01$.
In this work, for any summary statistic $S$, we calculate our derivatives as:

\begin{equation}
    \frac{dS}{d\Omega_m}=\frac{\langle S(\Omega_{m,fid}+\Delta\Omega_m)-S(\Omega_{m,fid}-\Delta\Omega_m)\rangle}{2\Delta\Omega_m},
\end{equation}
where the average is done over the $100$ realizations of these maps.
Each pair of maps with $\pm\Delta\Omega_m$ shares the same initial seeding, leading to low shot-noise levels due to cancellation.
We will denote this derivative estimate as $S'$.

Alternatively, we could have defined the derivative as:

\begin{equation}
    \frac{dS}{d\Omega_m}=\frac{\langle S(\Omega_{m,fid})\rangle -\langle S(\Omega_{m,fid}-\Delta\Omega_m)\rangle}{\Delta\Omega_m},
\end{equation}
where the averages are now over each individual set of realizations, since these maps do not share the initial seed.
We will denote this derivative estimate as $\tilde{S}'$.

If our value of $\Delta\Omega_m$ is small enough, these two derivative estimates should agree to within statistical uncertainty.
For statistical uncertainty, we approximate $S'$ as being noiseless due to the shot noise cancellations coming from identical initial seeding.
For $\tilde{S}'$, the majority of the shot noise will come from $\langle S(\Omega_{m,fid}-\Delta\Omega_m)\rangle$ since there are only $100$ realizations for this average, compared to the $1000$ fiducial realizations.
We therefore model all noise as arising from this term and assume that the maps are sampled from the fiducial covariance.
Therefore, $\Delta\Omega_m(S'-\tilde{S}')$ should vanish to within the noise of $\langle S(\Omega_{m,fid}-\Delta\Omega_m)\rangle$.
To test this, we plot histograms of:

\begin{equation}\label{eq:der_disc}
    \Delta\Omega_m(S'-\tilde{S}')/(\sigma(S)/\sqrt{100}),
\end{equation}
where $\sigma(S)$ is the elementwise standard deviation of the summary statistic estimated from our fiducial realizations.
The factor of $\sqrt{100}$ accounts for the majority of the noise coming from $\langle S(\Omega_{m,fid}-\Delta\Omega_m)\rangle$, which is an average of $100$ realizations.
If the summary statistics have no cross correlations, Eq.\ \ref{eq:der_disc} values should follow a Gaussian distribution.
To account for cross correlations, we sample $100$ vectors from the covariance matrix of the fiducial realizations with mean zero and average them elementwise.
We will call this averaged sample vector $D$ and plot out histograms of:

\begin{equation}\label{eq:der_disc_sample}
    D/(\sigma(S)/\sqrt{100}).
\end{equation}

\begin{figure*}
    \centering
  \includegraphics[width=0.90\textwidth]{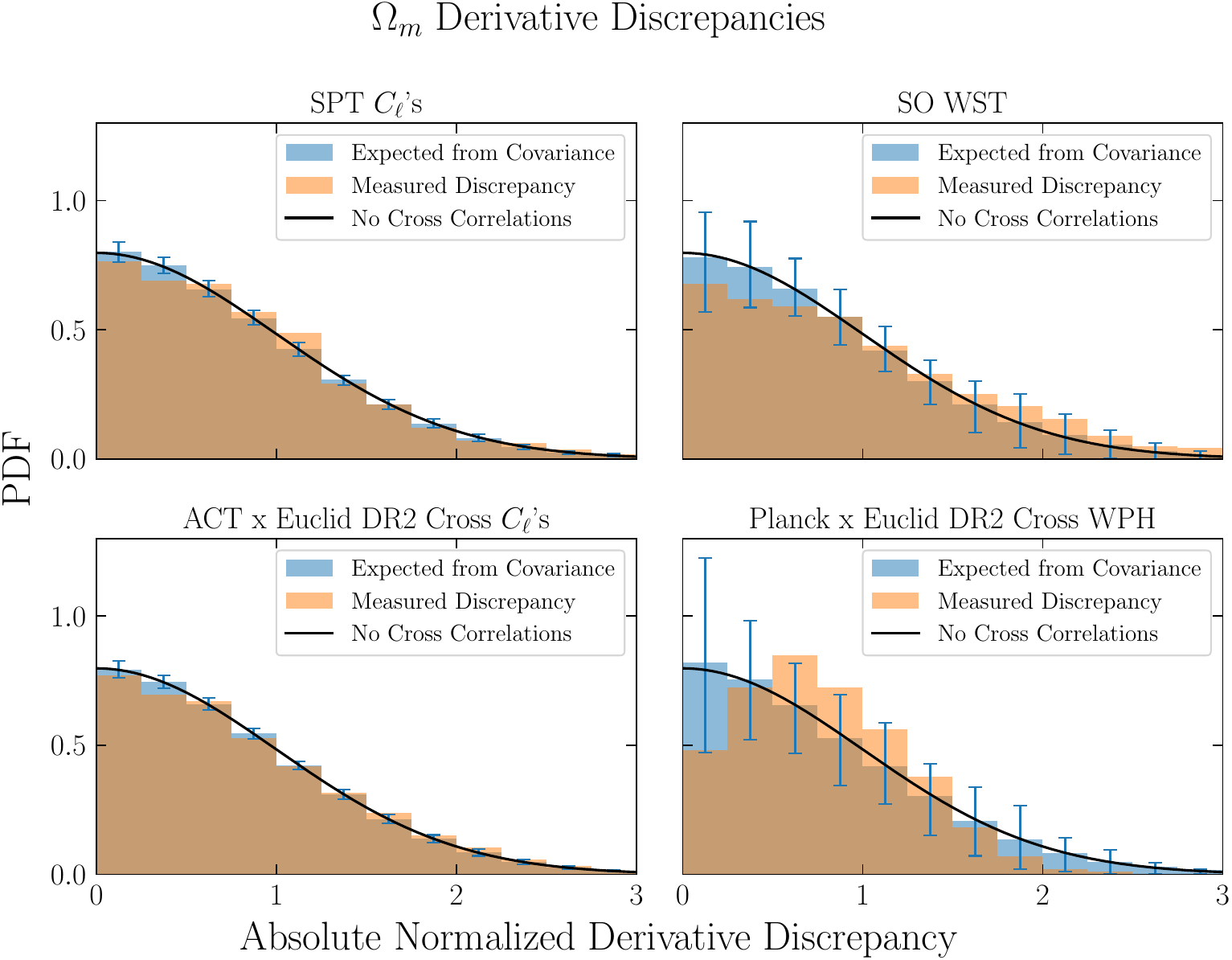}
  \caption{Distributions of normalized derivative discrepancies with $68\%$ error bars on histograms sampled from the fiducial covariance. One survey is picked at random for each summary statistic for brevity, but results for other surveys are comparable. Additionally, only $\Omega_m$ derivatives are shown for brevity, but $\sigma_8$ derivatives converge to comparable levels.}\label{fig:derivative_convergence}
\end{figure*}

If our derivative estimates are consistent with each other, the distributions from Eq.\ \ref{eq:der_disc} and Eq.\ \ref{eq:der_disc_sample} should match.
We show examples of the absolute values of these histograms with $68\%$ error bars on the sampled histogram in Figure \ref{fig:derivative_convergence}.
We find no significant differences between histograms for any summary statistic across surveys, so we conclude that our derivative estimates have sufficiently converged.

\clearpage

\section{Learned Binning Details}
\label{app:bin}
% \FloatBarrier

In order to properly construct a binning matrix, the function $f(x):[1,M]\rightarrow [1,N]$ defined in Section \ref{sec:analysis} must satisfy certain conditions.
Since the first entry of the summary statistic is mapped to the first bin, we must have $f(1)=1$.
Similarly, requiring that the last entry of the summary statistic is mapped to the last bin gives the requirement $f(M)=N$.
Bins should consist of entries in the summary statistic that are adjacent to one another (no gaps in the bins), which requires that $f(x)$ be a non-decreasing function. 
Finally, we want to ensure there are no empty bins.
There are numerous constraints one could impose on $f(x)$ to guarantee this property, and in this work we choose to require $f'(x)\leq1$.

To construct a binning matrix, we would generally have the $m^\text{th}$ column of the binning matrix be nonzero (and equal to $1$) only at the integer index closest to $f(m)$.
In this work, we slightly generalize binning to allow for one summary statistic entry to be mapped into two adjacent bins.
For example, if $f(2)=3.4$, the $2^\text{nd}$ summary statistic entry would be divided $60\%$ into bin $3$ and $40\%$ into bin $4$.
This splitting ensures that the determinant of the Fisher matrix is a continuous function of pivot positions.

As mentioned in Section \ref{sec:analysis}, our function $f$ is a linear interpolation between pivots, so that we perform optimization in a low-dimensional space.
Once we have picked a number of pivots, our loss function is the negative logarithm of the determinant of the Fisher matrix.
We include a prior term in our loss function to ensure that the pivots satisfy the conditions explained earlier in this section.
We optimize this loss function using the differential evolution method in SciPy \citep{scipy} with a population size of $20$.
The population is initially randomly seeded throughout the pivot space.

\section{Gaussianity of the Likelihood}
\label{app:gauss}
%\FloatBarrier

The Fisher forecast methods we use rely on our summary statistics, which follow a Gaussian distribution. 
With enough binning, the central limit theorem will ensure that our statistics are indeed Gaussian.
We test whether this is the case by producing a $\chi^2$ distribution of the statistics as in \citet{Valogiannis_2024}.
Specifically, if a summary statistic on any given realization is given by $\text{\bf O}_i$ with an average value of $\bar{\text{\bf O}}$ and covariance matrix $C$, the $\chi^2$ values can be given by:

\begin{equation}
    \chi^2 = (\text{\bf O}_i - \bar{\text{\bf O}})^T C^{-1}(\text{\bf O}_i - \bar{\text{\bf O}}).
\end{equation}

\begin{figure*}
    \centering
  \includegraphics[width=0.90\textwidth]{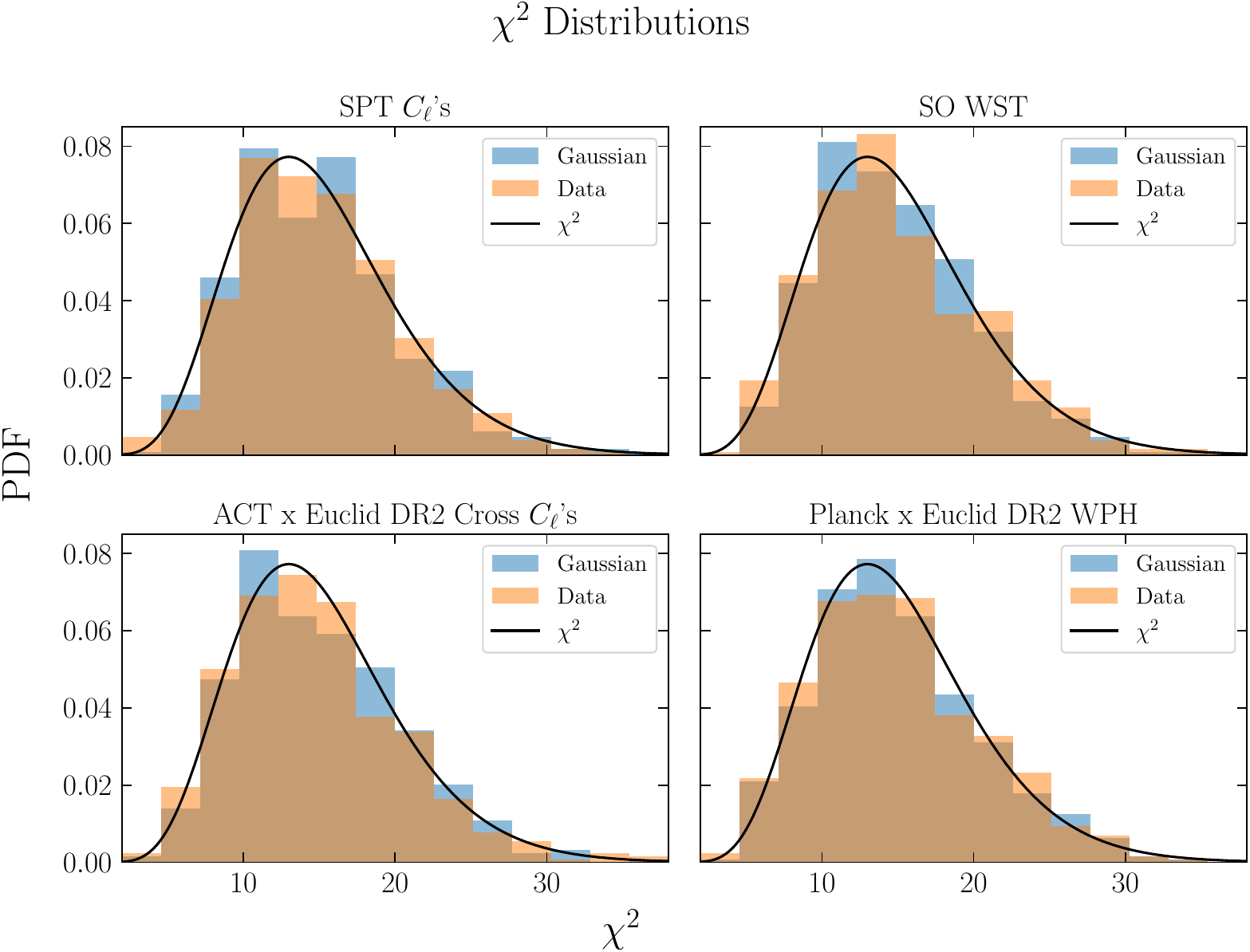}
  \caption{Probability density functions of the $\chi^2$ distribution for all summary statistics used. One survey is picked at random for each summary statistic for brevity, but results for other surveys are comparable. $500$ fiducial realizations (orange) are used for the distributions after being binned with a binning learned by the other $500$ fiducial simulations. Alongside the data distribution, we show the theoretical $\chi^2$ distribution with $15$ degrees of freedom (black) and samples drawn from a Gaussian distribution with the same mean and covariance as the data (blue). We find no evidence of significant departures from Gaussianity for any of our statistics.}\label{fig:chi2}
\end{figure*}

We perform this test with our binned data.
Specifically, we use $500$ fiducial realizations to learn the binning, then apply that binning to the remaining $500$ realizations to obtain $\chi^2$ values.
Since we use $15$ bins, if our statistics are Gaussian distributed, the calculated $\chi^2$ values should follow the theoretical $\chi^2$ distribution with $15$ degrees of freedom.
We verify this in Figure \ref{fig:chi2} where we find no significant differences between the $\chi^2$ distribution of our data, the theoretical $\chi^2$ distribution, and a Gaussian distribution from samples randomly generated with the same mean and covariance as our data.

\clearpage

\bibliographystyle{apsrev4-2_16.bst}
\bibliography{main}

\end{document}